\documentclass[manuscript,sigplan,twocolumn,nonacm]{acmart}

\usepackage{xspace}
\usepackage{subcaption}
\usepackage{tabularx}
\usepackage{makecell}

\usepackage{xcolor}
\usepackage{listings}
\usepackage[ruled,vlined,linesnumbered]{algorithm2e}

\definecolor{codekeyword}{RGB}{40, 80, 160}
\definecolor{codedecl}{RGB}{130, 50, 130}
\definecolor{codecomment}{RGB}{80, 120, 80}
\definecolor{codestring}{RGB}{160, 80, 0}
\lstdefinelanguage{wit}{
    morekeywords={
        package,use,func,true,false,bool
    },
    morekeywords=[2]={
        interface
    },
    morecomment=[l]{///},
    morecomment=[l]{//},
    morestring=[b]{`},
    sensitive=true,
    alsoletter={-,:,@,.},
}

\lstdefinelanguage{rust}{
    morekeywords={
        as,break,const,continue,crate,else,enum,extern,false,fn,for,
        if,impl,in,let,loop,match,mod,move,mut,pub,ref,return,self,
        Self,static,struct,super,trait,true,type,unsafe,use,where,while,
        async,await,dyn
    },
    morekeywords=[2]={
        Option,Result,Some,None,Ok,Err,String,Vec,Box
    },
    morecomment=[l]{//},
    morecomment=[s]{/*}{*/},
    morestring=[b]{"},
    sensitive=true,
}

\lstdefinelanguage{yaml}{
    morekeywords={
        rules,before,interface,inject
    },
    morecomment=[l]{\#},
    morestring=[b]{"},
    morestring=[b]{'},
    sensitive=true,
    alsoletter={-,:,*},
}

\AtBeginDocument{%
    }

\begin{document}


\newcommand{\ourtool}{\texttt{Splicer}\xspace}
\newcommand{\wirm}{\texttt{wirm}\xspace}
\newcommand{\cviz}{\texttt{cviz}\xspace}

\title[Write Once, Instrument Anywhere]{Write Once, Instrument Anywhere: Portable, Sandboxed Tooling with WebAssembly Components}

\author{Elizabeth Gilbert}
\orcid{0009-0005-7242-6179}
\affiliation{
    \institution{Carnegie Mellon University}
    \city{Pittsburgh}
    \state{Pennsylvania}
    \country{USA}
}
\author{Chris Fallin}
\orcid{0000-0002-6733-1803}
\affiliation{
    \institution{F5}
    \city{San Jose}
    \state{California}
    \country{USA}
}
\author{Heather Miller}
\orcid{0000-0002-2059-5406}
\affiliation{
    \institution{Carnegie Mellon University}
    \city{Pittsburgh}
    \state{Pennsylvania}
    \country{USA}
}
\author{Ben L. Titzer}
\orcid{0000-0002-9690-2089}
\affiliation{
    \institution{Carnegie Mellon University}
    \city{Pittsburgh}
    \state{Pennsylvania}
    \country{USA}
}


\begin{abstract}
  Monitoring and debugging deployed software is crucial for scalability, performance, and robustness.
  Cross-cutting concerns are pervasive across software ecosystems, yet engineers must address them with a fragmented toolbox.
  While high-level frameworks exist to offer application-level types and semantics, they often lock a solution into specific languages, runtimes, or hardware architectures.
  Portability often forces engineers to choose lower-level instrumentation mechanisms.
  For example, in-language frameworks can trace a request across processes, but multi-language contexts require instrumentation at the boundary, for example using eBPF.
  While these frameworks provide powerful capabilities, they often result in
  1) a tradeoff between application-level types or low-level bytes and packets,
  2) non-composable domain-specific tools,
  3) feature fragmentation across frameworks,
  and 4) tools that often run with unrestricted access to the program and its environment.
  
  We address these problems with \ourtool, a WebAssembly component interposition framework that
  1) provides semantically rich application-level types,
  2) enables pluggable tooling with powerful capabilities,
  3) reduces ecosystem fragmentation through porting across languages and hardware,
  and 4) confines each tool to a sandbox with explicitly granted permissions.
  We evaluate \ourtool on a suite of benchmarks demonstrating the pluggability of non-trivial tooling written in multiple source languages and their performance.
\end{abstract}

\maketitle

\section{Introduction}\label{sec:introduction}

Applications deployed in production often require cross-cutting monitoring and debugging, which is crucial for scalability, performance, and robustness.
As these fundamental needs are pervasive across software domains, many observability frameworks have arisen.
Common use cases are telemetry collection, security policy enforcement, payload enrichment, and optimization techniques.
To span a diverse set of use cases, we use the term \emph{tool} to refer to an implementation of a cross-cutting concern for a software system, similar to an \emph{aspect} in aspect-oriented programming~\cite{OrigAOP}.
Example tools are security policy enforcement, payload enhancement, profiling, tracing, and telemetry.

Systems like the OS kernel and profilers can provide observability without inserting instrumentation into target programs, but are often limited to binary data, network packets, and low-level events like page faults and system calls~\cite{ptrace, strace, DTrace, eBPF, etw, IntelPT}.
To access higher-level information, instrumentation is inserted into programs at the source, bytecode/IR, or binary level~\cite{OrigAOP,Oron,EEL,Pin,DynamoRIO}.
However, these approaches invasively inject code into programs, potentially altering their behavior inadvertently and are constrained to their specific language, runtime, and/or hardware.
On the other hand, VMs and emulators can keep programs and instrumentation separate, often imposing high overhead~\cite{QEMU,Valgrind}, which optimizations can reduce~\cite{Pin,DynamoRIO,wizard}.

This landscape poses a stark tradeoff for tools: application-level semantics \emph{or} non-invasive portability at the perimeter, but not both.
Our work offers a new, superior resolution of this tradeoff at a boundary that itself carries the application-level types.
Interposing at such a boundary is semantically rich, non-invasive, and ports across languages.

In this paper we present \ourtool, a framework that provides non-invasive, language-portable, and semantically rich interposition.
\ourtool supplies the missing instrumentation framework for the WebAssembly Component Model, a standard that makes rich, application-level types observable component boundaries.
Leveraging this rich information provided by Wasm components, \ourtool enables heterogeneous tooling (Table~\ref{tab:builtins}) with high performance (Figure~\ref{fig:builtins}).
We conduct a qualitative comparison with related frameworks in Table~\ref{tab:feature-matrix} showing that \ourtool is the only framework with language portability, application-level typing semantics, and high-granularity without extra supporting infrastructure like agent processes.
To empirically evaluate the overhead of \ourtool in both runtime and code size, we conduct a series of experiments that show it imposes acceptable overhead on both microbenchmarks and real-world applications.

\subsection{WebAssembly (Wasm)}\label{subsec:wasm}
WebAssembly~\cite{WasmPldi} is a portable bytecode format targetable by many source languages including C, C++, Rust, Go, Scala, OCaml, and many more.
It was originally introduced on the web, but has since gained traction in other domains including embedded systems~\cite{WasmIndMach, wasm-iot, wpu}, cloud computing~\cite{CloudflareWasm, wasmcloud-cncf}, and edge computing~\cite{FastlyEdge, WasmSD-Edge} due to its sandboxed execution environment, strict execution model, polyglot capability, near-native performance, and high portability.

When a program is compiled to Wasm, a \emph{core module} is generated that contains the program's logic that can \emph{export} functionality to external callers.
Values that can cross this boundary are limited to core Wasm types such as \texttt{i32}, \texttt{f32}, etc.
Recently, the W3C WebAssembly Community Group and the Bytecode Alliance have introduced the Component Model~\cite{component-model} as a module composition framework, described in depth in Section~\ref{sec:background}.
While instrumentation frameworks~\cite{brewasm,AspectWasm,Walrus,whamm} have emerged for the the core Wasm ecosystem, none to date support components.

\begin{figure}
    \includegraphics[width=1.00\columnwidth]{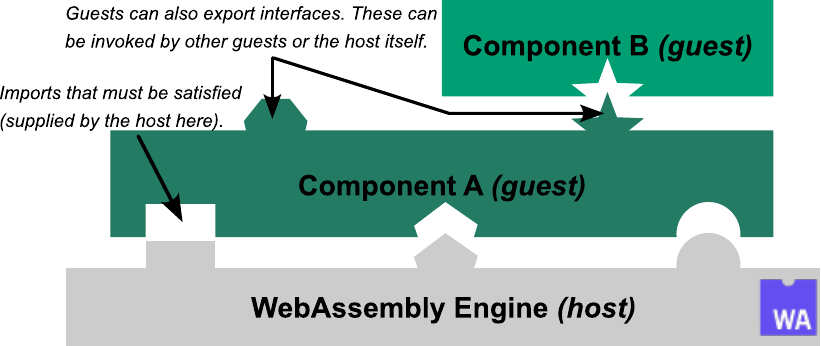}
    \caption{
        A high-level view of a Wasm \emph{component}.
        Components can import and export \emph{interfaces}, comprising its \emph{world}.
        Type compatibility is shown here as complementary shapes.
    }
    \label{fig:component}
\end{figure}

In the Component Model, a \emph{core module} is encapsulated by a Wasm \emph{component} that exposes richly typed interfaces defined by the Wasm Interface Types (WIT) interface description language (IDL).
This enables the \emph{exported} functionality of a component to carry rich semantic meaning as the values flowing across component boundaries use WIT-level types (strings, lists, records, etc.).
Through \emph{composing} components with a Wasm Composition (WAC) script, components produced by different source language toolchains can interoperate safely in-process, shown in Figure~\ref{fig:component}.
Creating such a \emph{composition} from a set of $N$ Wasm components produces a single Wasm component binary containing the full set components with explicitly-defined communication topologies over their well-typed interfaces.
So, while the composition executes as a single process, it does not need to be viewed as a monolithic unit.
Static analysis can reconstruct its logical subunits with application-level typing semantics of their communication.
These well-typed interfaces present a natural pointcut for interposition and is the approach used by \ourtool to configure tooling placement.

\subsection{Contributions}\label{subsec:contributions}
In this paper we outline the design of \ourtool, a framework on which tools can be built once and applied portably across languages.
Our contributions include:
\begin{itemize}
    \item \textbf{The first interposition framework for Wasm components.}
      Prior work has only provided instrumentation for core Wasm modules, making use cases like security policy enforcement, payload enhancement, and proxying difficult, which typically require higher level information.
      
    \item \textbf{A new approach for preserving semantically-rich types during interposition.}
      Tools can be implemented with minimum data access and \ourtool optimally places them using language-portable semantics.
          
    \item \textbf{Increased pluggability with automatic tool adaptation.}
      Tools can be written against a single contract and \ourtool \emph{adapts} them to any interface and renders payloads in a generic view.




    \item \textbf{A suite of tooling demonstrating \ourtool's expressiveness.}
          We implement a set of common, non-trivial instrumentation tasks, shown in Table~\ref{tab:builtins}.

    \item \textbf{A Wasm component port of the standard DeathStarBench Hotel Reservation benchmark.}
          We ported this well-known benchmark to enable comparison between the native and componentized deployments.

    \item \textbf{An empirical evaluation across microbenchmarks and a standard microservice benchmark.}
          We show that the Component Model's safe in-process interoperability lets \ourtool instrument communication with high throughput.

\end{itemize}

\section{Background}\label{sec:background}

The WebAssembly Component Model~\cite{component-model}, offers a programming standard that facilitates safe interoperability through well-typed interfaces defined by the Wasm Interface Types (WIT)~\cite{wit} Interface Description Language (IDL).
An \emph{interface} is a collection of functions and their corresponding type definitions.
The term \emph{world} refers to a collection of interfaces and types that a component depends on (\emph{imports}) and/or provides (\emph{exports}) at its outermost layer, shown in Figure~\ref{fig:component}.

\subsection{Component Composition}\label{subsec:composition}
A component is a static definition; instantiating it yields an \emph{instance} with its own state.
\emph{Compositions} link instances together so they interoperate within a single process.
The format of a composition is a directed graph of connected instances, creating a topology of their communication.
A composition is produced by running a WebAssembly Compositions (WAC)~\cite{wac} script that explicitly expresses how to link instances.
For example: ``Instance A's \emph{imported} dependency is satisfied by instance B's complementary \emph{export}.''
At runtime, when A calls this \emph{imported} interface, it will invoke the interface from B wired through in the WAC script.

\subsection{Communication Over WIT}\label{subsec:communication}
A core Wasm module contains the actual program logic, while its encapsulating component defines its incoming and outgoing interfaces using \emph{WIT}'s~\cite{wit} type system.
The \emph{canonical ABI}~\cite{canon-abi} translates between high-level WIT types (strings, lists, etc.) and low-level core Wasm types (i32, f32, etc).
Components are \emph{shared-nothing}; one component cannot access any other component's state, but must communicate via pass-by-copy semantics.
A cross-component call \emph{lifts} payloads from core-Wasm to WIT-level types, then \emph{lowers} them across the component boundary.
Lifting is not a runtime step, it is the semantic contract between caller and callee that the engine uses to validate the composition and generate \emph{lowering} logic that performs the payload copy.

\begin{figure*}
    \includegraphics[width=1.00\linewidth]{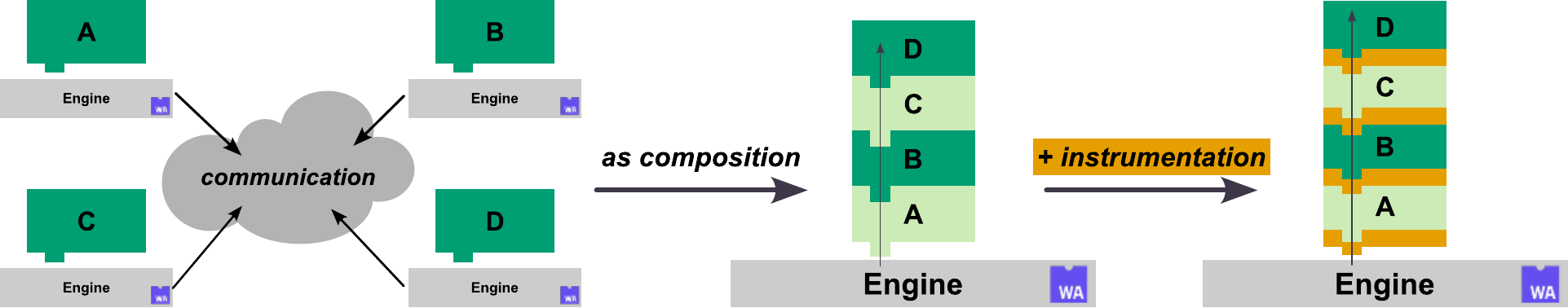}
    \caption{
        Wasm programs can communicate over arbitrary protocols (\emph{left}).
        A \emph{composition} can instead do this safely, in-process (\emph{middle}).
        This in-process communication is invisible at the program's perimeter (\emph{right}), so instrumentation must be interposed.
    }
    \label{fig:sub-comp-instr}
\end{figure*}

\begin{table*}
    \centering
    \small
    \begin{tabular}{|l|l|l|l|l|l|l|}
        \noalign{\hrule height 1.5pt}
        & \textbf{Isolation} & \textbf{On failure} & \textbf{Types}
        & \textbf{Cross-Language} & \textbf{Versioning} \\
        \noalign{\hrule height 1.5pt}
        \hline
        \textbf{dlopen}
        & none & corrupts caller & structs, raw pointers
        & same ABI & fragile \\
        \hline
        \textbf{C FFI}
        & none & corrupts caller & bytes, raw pointers
        & any (C ABI) & fragile \\
        \hline
        \textbf{Wasm Module}
        & sandboxed & contained (trap) & i32/i64/f32/f64/etc
        & manual glue code & fragile \\
        \hline
        \textbf{Wasm Component}
        & sandboxed & contained (trap) & strings, records,\ldots
        & safe, typed & structural \\
        \hline
        \textbf{IPC (pipe/socket)}
        & separate process & partial failure & raw bytes, no framing
        & needs a codec & schema \\
        \hline
        \textbf{RPC / gRPC}
        & separate process & partial failure & messages, schema types
        & safe, typed & schema \\
        \hline
    \end{tabular}
    \caption{
        A comparison of interoperability mechanisms.
        The Component Model alone is safe, in-process, and typed.
    }
    \label{tab:comm}
\end{table*}

Table~\ref{tab:comm} shows some examples of polyglot communication mechanisms.
While languages can interoperate in-process via \texttt{dlopen}, the C FFI, and core Wasm modules, doing so requires low-level typing and unsafely passing raw pointers to shared memory.
Inter-process communication (IPC) enables isolated processes to interoperate through well-defined protocols, increasing safety.
However, communication across processes or networks has high overhead and complex failure semantics.
The Component Model facilitates safe, in-process, and performant language interoperability while retaining application-level types, as shown in Figure~\ref{fig:sub-comp-instr}.

\subsection{Component Failure Semantics}\label{subsec:failure}
The Component Model has favorable failure semantics because sandboxing isolates faults and composition makes call completion explicit.
WIT provides a \texttt{result} type to gracefully handle faults; however, \emph{unexpected} faults still trap.
On a trap, the \emph{shared-nothing} property guarantees that a caller's memory cannot be corrupted.
Composition avoids the \emph{partial-failure} semantics of distributed communication.
For example, RPC separates the request and response into two messages passed over a channel which can fail delivery.
A client that sees no response cannot tell if the peer ran.
In contrast, a cross-component call has no intervening channel, it happens synchronously, suspending the caller until the callee returns with a \texttt{result} error or it traps.
Async calls relax this, resulting in the possibility of \emph{partial completion}, requiring the caller to appropriately handle the failure modes of in-process concurrency.
Such communication avoids reasoning about unreachable or independently failing peers possibly distributed across a network.

\subsection{Resources}\label{subsec:intro-resources}

The Wasm Component Model offers the \texttt{resource} type to define objects.
They are useful for defining capabilities such as database connections and file descriptors.
Rather than copying entire object instances across component boundaries, resources are passed as opaque IDs, called \emph{handles}.
The component that \emph{implements} a resource contains all instances of that resource and its methods.
When a method is invoked on a \texttt{resource}, it performs a cross-component call to its \emph{implementing component} which uses the handle to look up the instance and invoke the target method.


\section{The \ourtool Framework}\label{sec:splicer}
\ourtool is a framework that interposes Wasm components into arbitrary compositions, enabling language-portable tooling with application-level semantics.
The framework is provided an \emph{interposition plan} containing a list of \emph{rules} that express where tools should be placed in a composition.
To execute the plan, \ourtool first discovers a component's composition, then applies the plan's rules to place tools within it.
Finally, it uses WAC to emit the instrumented composition as a single component binary.

\subsection{Discovering the Composition}\label{subsec:discovery}
\ourtool discovers a composition graph directly from a component binary, without access to source code, a WAC script, or WIT definitions.
Doing so requires careful handling of the component's internals.
In Wasm's binary format, items are referenced by integer indices into an \emph{index space}.
A component has 13 such index spaces, all of which must be tracked for correct reference resolution.
Components can also nest to arbitrary depth, each containing their own set of index spaces, and a nested component can reference items in any enclosing component's spaces.
We capture this in a \texttt{Visitor}-pattern API with precise node resolution, made available in the open-source \wirm Rust crate\footnote{
    https://crates.io/crates/wirm 
}.

Instantiating a Wasm component requires satisfying its \emph{imports} by providing complimentary instance \emph{exports}.
For example, in Figure~\ref{fig:component}, B's star-shaped import is satisfied by A's star-shaped export.
Such dependencies are expressed in the bytecode format as arguments on instantiation blocks.
Consequently, the discovery algorithm iterates over these blocks to identify instances, which become \emph{nodes}, and uses block arguments to identify interface dependencies between instances, which become directed \emph{edges}.
Any dependencies not supplied as arguments must be satisfied by the host.
To discover the type signature of an edge, type references are recursively resolved to their concrete type.
The outermost component layer is also analyzed to identify its exports to discover its \emph{public surface}.
Our algorithm implementation has been factored into the open-source \cviz\footnote{
    https://crates.io/crates/cviz 
} Rust crate.

\subsection{The Interposition Plan}\label{subsec:config}

Users provide an interposition plan as a YAML file containing a list of \emph{rules} that specify \emph{where} to interpose tools.
For large compositions, the interposition plan side-effects are cumbersome to track, so we leverage the \cviz Rust crate to render a visual preview of the resulting composition.
Interposed tools are shown as new, highlighted nodes with tags that point to the rule that configured them.

\subsubsection{\emph{Where} to Inject}\label{subsubsec:where}
\ourtool has expressive matching capabilities to find interposition points in a composition.
Graph nodes carry their instance names from the producing WAC script while edges carry the name of their WIT interface.
Interface names have the following convention: ``\emph{namespace:package/interface@semver}''.
Users can leverage \textbf{wildcards} to match against a set of node and edge names.

\textbf{Edge/Interface ``shapes''} are exposed to constrain matches against function properties.
We currently apply constraints across all functions in an interface edge, they must all meet the constraint to match.
Properties include the following, but can be extended as new use-cases arise:
\begin{itemize}
    \item \emph{async}: whether functions are async,
    \item \emph{scope}: where functions live; as interface-level free functions or as resource methods,
    \item \emph{args} and \emph{results}: generic constraints on payload types.
\end{itemize}
Generic constraints enable tools to target type-compatible interfaces.
For example, a tool that conditionally skips the downstream may only work when results are \texttt{defaultable} (like a string) and \texttt{concrete} (non-resource).

The \texttt{before} rule interposes on a node's incoming edge, while the \texttt{between} rule interposes on a connecting edge of two nodes.
The \texttt{on\_node} rule interposes on any edge of a specified node.
It can be constrained on edge direction: \texttt{inbound}, \texttt{outbound}, or \texttt{both}.
The \texttt{on\_subgraph} rule treats a set of nodes as a single unit.
For a group of nodes to qualify as a subgraph, every node must be reachable by every other node without following paths through nodes outside the set (the edges must be between nodes both inside the set).
If the subgraph is found, it is treated as if it were a single node with its own incoming and outgoing edges.
For example, with subgraph \texttt{[b, c]} and edge direction as ``both,'' the graph $a \rightarrow b \rightarrow c \rightarrow d$ becomes $a \rightarrow T \rightarrow b \rightarrow c \rightarrow T \rightarrow d$ after interposing tool \texttt{T}.

\subsubsection{\emph{What} to Inject}

\begin{table*}
    \centering
    \small
    \begin{tabular}{|l|l|l|l|}
        \noalign{\hrule height 1.5pt}
        \textbf{Name} & \textbf{Tier} & \textbf{LoC by Language} & \textbf{Description} \\
        \noalign{\hrule height 1.5pt}

        \textbf{hello-tier1} & 1 & Rust 32; Go 17 & Prints a greeting before/after each call. \\
        \hline
        \textbf{otel-bare-logs} & 1 & Rust 101; Go 128 & OTel log on each call's return. \\
        \hline
        \textbf{otel-bare-spans} & 1 & Rust 147; JS 106 & OpenTelemetry spans around calls. \\
        \hline
        \textbf{otel-bare-metrics} & 1 & Rust 190; JS 150 & OTel call-count + duration metrics. \\
        \noalign{\hrule height 1pt}

        \textbf{hello-tier2} & 2 & Rust 48; Go 43 & Prints a greeting and args/results before/after each call. \\
        \hline
        \textbf{recorder} & 2 & Rust 137; Go 156 & Record every wrapped call's args + results. \\
        \hline
        \textbf{otel-payload-metrics} & 2 & Rust 482; JS 307 & Derive OTel metrics (size, shape, etc.) about call payloads. \\
        \noalign{\hrule height 1pt}

        \textbf{hello-tier3} & 3 & Rust 30 & Prints a greeting before/after each call. \\
        \hline
        \textbf{redact-strings} & 3 & Rust 77 & Rewrite string content in payloads matching regexes. \\
        \hline
        \textbf{memoize} & 3 & Rust 135 & Return cached result on repeat calls, skipping the downstream. \\
        \noalign{\hrule height 1pt}

        \textbf{hello-tier4} & 4 & Rust 30 & Prints a greeting before returning a default result. \\
        \hline
        \textbf{chaos-err} & 4 & Rust 31 & Returns a freshly-sampled Err for every wrapped call. \\
        \hline
        \textbf{replayer} & 4 & Rust 54 & Returns each wrapped call's result from a recorded trace. \\
        \hline
        \textbf{fuzz-results} & 4 & Rust 319 & Returns a random value for every call. \\
        \hline
    \end{tabular}
    \caption{
        \ourtool's suite of configurable tools for common instrumentation tasks.
        Tiers 1--2 support any component-targetable language; tiers 3--4 are currently Rust-only.
        LoC counts tool logic only, excluding blank lines, comments, and plumbing (e.g., payload serialization) that \ourtool factors into a Rust SDK.
        Equivalent SDKs for other languages would be a one-time cost.
    }
    \label{tab:builtins}
\end{table*}

The list of tools to inject at match points is provided in runtime-execution order.
\ourtool provides configurable \emph{builtins} to support common instrumentation tasks, see Table~\ref{tab:builtins} for the full list.


\subsection{Enabling Interface-Portable Tooling}\label{subsec:adapters}

\begin{figure*}
    \includegraphics[width=1.00\linewidth]{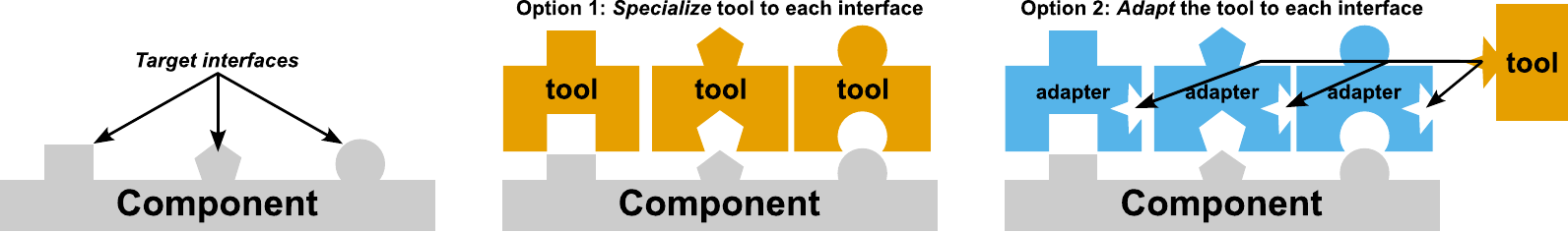}
    \caption{
        A component can export $N$ unique interfaces a tool must target.
        The tool can either shipp as $N$ specialized variants, \emph{or} rely on \ourtool's adapter to target all $N$ automatically.
    }
    \label{fig:adapters}
\end{figure*}
\begin{figure*}
    \includegraphics[width=1.00\linewidth]{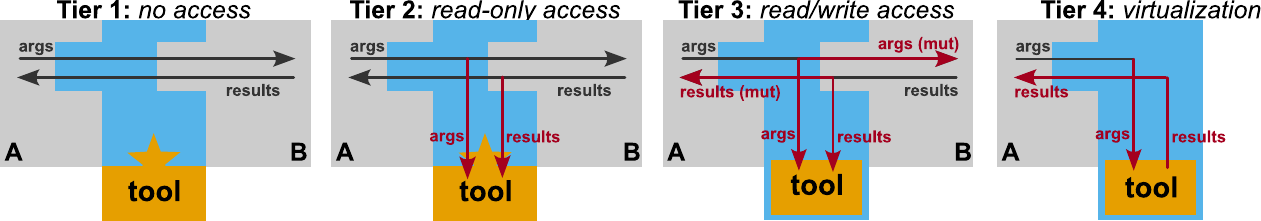}
    \caption{
        \ourtool generates adapters exposing different access levels based on a tool's needs.
        Tiers 1--3 differ in payload access, while tier 4 enables complete virtualization of a downstream dependency.
    }
    \label{fig:tiers}
\end{figure*}

Figure~\ref{fig:adapters} shows how \ourtool supports \emph{specialized}, interface-compatible tooling and tooling that targets \ourtool's generic interface, requiring adaptation.
\ourtool offers multiple adapter types, named \emph{tiers}, which differ in data access privilege and control over the downstream, shown in Figure~\ref{fig:tiers}.
In a landscape where an increasing percentage of new code is generated by AI, structural guardrails are essential to improve system safety and security.
So, while every tier's capabilities are achievable by tier 3, we intentionally created a capability ``ladder'' to enable users to write tools with minimum access.

\subsubsection{The Adapter Design.}
Due to design tradeoffs, users express tooling differently depending on the tier they target.

\textbf{Tiers 1 and 2 are provided as WIT worlds}, enabling tools to be implemented in any language that compiles to components, as Table~\ref{tab:builtins} shows.
Each tool lives in a sandboxed component, separate from the adapter, so it has no direct payload access.
\ourtool generates the adapter component's Wasm directly which contains a core module that invokes callbacks as cross-component calls.
These tiers therefore add at least two extra cross-component calls: one into the adapter and one per implemented callback.
Because tools ship as component binaries, users fully control any dependencies they introduce.
For example, a tool can propagate metadata across instances by writing to an external agent, provided the host supports such communication.


\textbf{Tiers 3 and 4 are provided as Rust \texttt{traits}.}
As these tiers allow payload mutation, sandboxing tools from the adapter is not necessary, so tooling logic is contained directly in the adapter.
Callbacks are then simple Rust \texttt{trait} invocations, so these tiers add only one extra cross-component call into the adapter.
We implement these adapters in Rust rather than hand-authoring them in Wasm because payload mutation is complex, especially for WIT resources, which require \ourtool to \emph{reflect} on handles.
Writing adapters in Rust lets us inherit a mature toolchain, to which we offload  (1) lowering payloads from their abstract representation back into the canonical ABI, (2) synthesizing the backing component for a WIT resource, and (3) tracking \texttt{own}/\texttt{borrow} resource semantics.
The design extends cleanly to new capabilities such as resource reflection.
Because \ourtool controls the adapter's WIT world, tools cannot introduce dependencies by adding imports, but we plan to address this in future work.

\subsubsection{Tier1: Name-Only Hook}\label{subsubsec:tier1}
Tier 1 tools have no data access but can reflect on the function being invoked, which is useful for tooling such as OpenTelemetry span generation.
Tools can implement any of the WIT function callbacks:
\begin{lstlisting}[language=wit, basicstyle=\footnotesize\ttfamily,label={lst:tier1}]
on-call: func(call: call-id);
on-return: func(call: call-id);
should-call: func(call: call-id) -> bool;
\end{lstlisting}
The \texttt{on-call} callback is invoked before the downstream, \texttt{on-return} is invoked after, and \texttt{should-call} can conditionally skip the downstream, requiring a \texttt{void} downstream function.
The \texttt{call-id} is a WIT \texttt{record} containing the interface and function names and a unique ID that can be used to correlate \texttt{on-call} and \texttt{on-return} invocations.
To make callback more efficient, \ourtool pre-allocates memory for the \texttt{call-id} ABI slots and prefills it with the statically-known interface and function names.
The unique ID is calculated and written to its respective slot at invocation time.

\subsubsection{Tier2: Read-Only Data}\label{subsubsec:tier2}
Like tier 1, tier 2 tools can reflect on the invoked function, but additionally have read-only access to its payloads.
Tools implement WIT callbacks:
\begin{lstlisting}[language=wit, basicstyle=\footnotesize\ttfamily,label={lst:tier2}]
on-call: func(call: call-id, args: list<field>);
on-return: func(call: call-id, result: option<field-tree>);
should-call: func(call: call-id, args: list<field>) -> bool;
\end{lstlisting}
Payloads are flattened into an arena/pointer representation:
\begin{lstlisting}[language=wit, basicstyle=\footnotesize\ttfamily,label={lst:tier2-types}]
variant cell {
    boolean(bool),
    list-of(list<u32>),   // child indices, in order
    record-of(u32),       // --> record-infos[idx]
    resource-handle(u32), // --> handle-infos[idx]
    ...
}
record record-info {
    type-name: string,
    /// (field-name, cell-index)
    fields: list<tuple<string, u32>>,
}
record handle-info {
    type-name: string,
    id: u64,
}
record field-tree {
    root: u32,
    cells: list<cell>,
    record-infos:  list<record-info>,
    handle-infos:  list<handle-info>,
    // ... other side tables
}
record field {
    name: string,
    tree: field-tree,
}
\end{lstlisting}
A \texttt{field-tree} is a single payload whose \texttt{root} points to the root \texttt{cell} in the \texttt{cells} arena.
Compound cells reference their children by indexing into the arena.
Type metadata, such as a \texttt{record}'s field names, is stored in \texttt{*-info} side tables so that every \texttt{cell} packs into $\le 8$ bytes.
This uniform width lets the adapter allocate and serialize the arena as one flat array, rather than chasing nested, variably-sized nodes.
Handles (resources, streams, and futures) are represented by their integer identifier, which cannot be used to invoke the handle's methods.
We strip those capabilities to prevent inadvertent mutation while letting tools correlate the same handle across calls.
The \texttt{field} type represents a named function argument, carrying the name from its WIT function signature.

Arguments are provided to a callback as a list of \texttt{field}.
Results are a single \texttt{field-tree} payload.
When the size of every payload in a target interface is known statically, \ourtool performs an optimization that pre-allocates memory regions and generates a side-table of pointers to them.

\subsubsection{Tier3: Read/Write Data}\label{subsubsec:tier3}
This tier provides read/write access to arguments and results.
Tools implement:
\begin{lstlisting}[language=rust, basicstyle=\footnotesize\ttfamily,label={lst:tier3}]
pub trait TransformStrategy<Args, R> {
    async fn handle(&self, call: CallId, args: Args,
        downstream: impl AsyncFn(Args) -> R,
    ) -> R;
}
\end{lstlisting}
Arguments are bundled in a generated struct, \texttt{Args}, containing parameter payloads as fields.
The \texttt{downstream} closure can be invoked by the tool to call the downstream component and get its results, \texttt{R}, which are simply the declared WIT type.
When an argument or result is a handle, the handle is actually passed to the tool.
In principle, the tool could invoke its methods, but first reflection must be provided to enable tools to discover a handle's methods.

\subsubsection{Tier4: Virtualize}\label{subsubsec:tier4}
This tier provides read/write data access, but no ability to invoke the original downstream.
In fact, the component-level downstream import is not generated in the adapter.
Tools implement:
\begin{lstlisting}[language=rust, basicstyle=\footnotesize\ttfamily,label={lst:tier4}]
pub trait VirtualizeStrategy<Args, R> {
    async fn handle(&self, call: CallId,
        args: Args) -> R;
}
\end{lstlisting}
This tier is very similar to tier 3's argument bundling and resource limitations due to a lack of reflection capabilities.
We plan to provide such features in future work.

\subsection{Interposing on WIT Resources}\label{subsec:resources}
Resources are represented as \emph{handles} passed across ABI boundaries with explicit ownership semantics.
A handle acts as a ``smart pointer'' to a resource instance, which lives in an \emph{implementing component} that holds the resource's state and methods.
Invoking a resource method sends the operation to this component, which uses the handle to look up the instance and runs the method in its own context.

To accomplish instrumenting a resource, \ourtool generates a resource that \emph{wraps} the target resource.
The \emph{wrapper}, of type \texttt{T'}, exposes a WIT world identical to the original resource's, of type \texttt{T}.
Through having an identical component surface, \texttt{T'} handles can be passed anywhere \texttt{T} handles are.
Whenever a method is invoked on \texttt{T'}, it runs the instrumentation callbacks, then forwards the call to the corresponding method on \texttt{T}.
If instrumentation should only be active within a composition \emph{subgraph}, translation between \texttt{T} and \texttt{T'} is necessary at subgraph boundaries.
\begin{figure}
    \includegraphics[width=0.80\columnwidth]{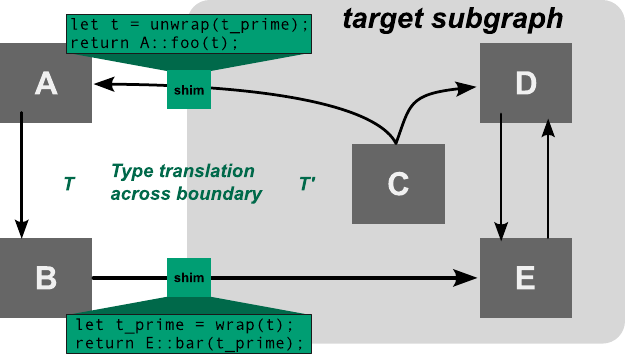}
    \caption{
        \ourtool wraps instrumented resources with a wrapper.
        If instrumentation is constrained to a subgraph, it must be wrapped/unwrapped at the boundary.
    }
    \label{fig:resource_wrappers}
\end{figure}
Figure ~\ref{fig:resource_wrappers} shows how small component \emph{shims} are interposed at subgraph boundaries to wrap the raw resource on incoming edges and unwrap on outgoing edges.
This way the resource exiting the subgraph is the original, unwrapped resource \texttt{T} while the resource within the subgraph carries instrumentation.
This approach cannot capture method calls on a resource when its handle is produced by a free function of an interface that the implementing component exports.
Interposing a resource's methods requires substituting its type, but that type is owned by the implementing component, and we cannot override its definition.
So once such a component hands out a handle, calls back onto it are invisible to us.
This is a consequence of the Component Model's handle semantics, where the resource's identity stays bound to its implementing component.

%

\subsection{The Expressiveness of \ourtool}\label{subsec:expressiveness}
\ourtool supports common instrumentation tasks.
\textbf{Matching all edges}, with a \texttt{before} rule and a \texttt{``*''} interface, enables use cases such as \emph{distributed tracing / span generation}~\cite{xtrace,Dapper}, \emph{data provenance}~\cite{data-prov}, and \emph{latency attribution}~\cite{TheMM}.
Data provenance requires context propagation, which is currently achievable in tiers 1--2, but not in tiers 3--4 which do not support additional host imports.
This is planned as future work.
\textbf{Matching by type shape}, by constraining to functions with certain typed payloads, enables \emph{resilience} patterns such as retry and timeout~\cite{resilience} and \emph{fault-injection testing}~\cite{Filibuster,Gremlin}.
\textbf{Matching by topology}, using the \texttt{on\_node} and \texttt{on\_subgraph} rules, enables use cases like \emph{record/replay}~\cite{replay} and \emph{mocking}~\cite{mocking} of a single node or an entire subgraph.

\section{Evaluation}\label{sec:eval}

We evaluate the costs of \ourtool across a suite of microbenchmarks and compare it to other frameworks on the DeathStarBench~\cite{DeathStarBench} Hotel Reservation benchmark.

\subsection{The Static Cost}\label{subsec:static-cost}

\begin{figure*}
    \centering
    \includegraphics[width=1.00\linewidth]{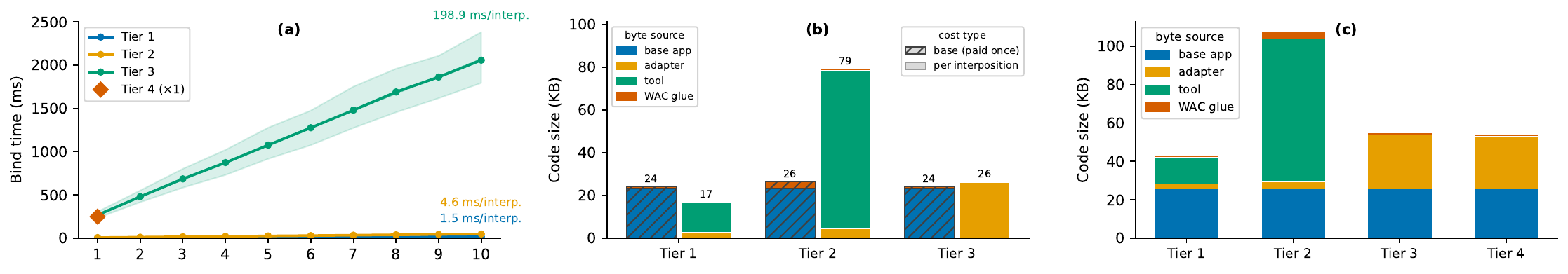}
    \caption{
        Static base cost of interposing a \emph{noop} tool.
        (a) Bind time as interpositions increase.
        (b) Code-size growth decomposed by source component bytes.
        (c) Total byte-source attribution of a single interposition across tiers.
    }
    \label{fig:build_time}
\end{figure*}

Figure~\ref{fig:build_time} shows static costs in time and space.
We measure the \emph{bind time}, which is the time to rewrite the application code, and code size of the instrumented application.
As tier 4 completely replaces a composition's subgraph, only a single interposition is meaningful, so we report its bind time in chart (a) and code size increase in chart (c) for one interposition.
Chart (a) shows the bind time as the number of interpositions grows.
Tiers 1--2 are fast because \ourtool generates Wasm directly, while tiers 3--4 incur a higher cost as \ourtool generates and compiles Rust code at bind time.
Chart (b) shows the code size breakdown.
We attribute code size to its source by measuring each binary before composing them into a single Wasm binary.
WAC glue is the difference between the resulting binary's size and the sum of all pre-existing binaries.
The left bar shows the \emph{base} cost and the right shows the increase per interposition.
Tier 3--4 adapters directly embed tool code and lowering logic, making them significantly larger than tier 1--2's.
The per-interposition code size increase reveals an optimization opportunity for WAC:
rather than copying byte-identical components once per instantiation, it could reuse the same component across instantiation sites.
This would reduce \ourtool's per-interposition cost to just the WAC glue of the new instantiation.
Chart (c) shows the code size breakdown for a single interposition, enabling direct comparison across all tiers.

\subsection{The Dynamic Cost}\label{subsec:dynamic-cost}
We evaluate \ourtool on a suite of microbenchmarks and compare it to state-of-the-art frameworks on the DeathStarBench~\cite{DeathStarBench} Hotel Reservation benchmark.
We ran experiments on a machine with two Xeon Platinum 8168 CPUs @ 2.70GHz with 96 total cores and 384GiB of DDR4 RAM.

\subsubsection{Base \ourtool Overhead}\label{subsubsec:perf-noop}

\begin{figure*}
    \includegraphics[width=1.00\linewidth]{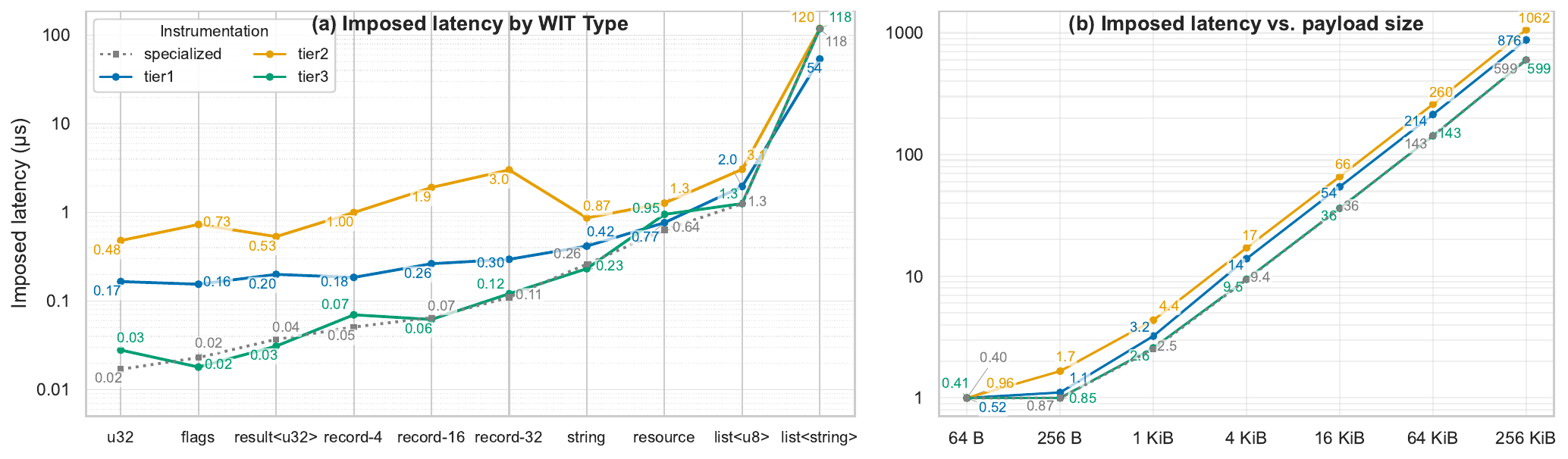}
    \caption{
        Base \ourtool overhead with \emph{noop} tools.
        (a) The imposed latency by WIT type, see Table~\ref{tab:wit-type},
        (b) Imposed latency as a \texttt{list\textless u8\textgreater}'s size increases.
    }
    \label{fig:base_overhead}
\end{figure*}

Figure~\ref{fig:base_overhead} shows the base overhead of \ourtool by measuring throughput with a \emph{noop} interposition.
The \emph{specialized} measurements show the effect of interposing a directly compatible tool on an interface.
The tier 1--3 measurements show the overhead of a \ourtool-adapted tool; tier 4 was omitted as it removes downstream calls.
The overhead is measured by subtracting the p50 latency of an uninstrumented component from the latency of a \emph{noop}-interposed composition.

\begin{table*}
    \small
    \begin{tabular}{|l|p{0.30\linewidth}|p{0.48\linewidth}|}
        \noalign{\hrule height 1.5pt}
        WIT Type    & Description                                & Marshalling Behavior                                                                                                \\
        \noalign{\hrule height 1.5pt}
        u32         & scalar                                     & in-register                                                                                                         \\
        \hline
        list-u8     & a list of bytes; length 1024               & memcopy                                                                                                             \\
        \hline
        string      & a list of utf8 bytes; length 1024          & memcopy + UTF-8 validation                                                                                          \\
        \hline
        record-4    & a record with 4 scalar fields              & as arg: flattened, in-register; as result: spilled, memcopy             \\
        \hline
        record-16   & a record with 16 scalar fields             & as arg: flattened, in-register; as result: spilled, memcopy                                                         \\
        \hline
        record-32   & a record with 32 scalar fields             & as arg: spilled, memcopy; as result: spilled, memcopy                      \\
        \hline
        list-string & a list of 1024 single-character strings    & memcopy + UTF-8 validation \emph{per-element}                                                      \\
        \hline
        result-u32  & variant                                    & as arg: flattened, in-register; as result: spilled, memcopy                                                         \\
        \hline
        flags-32    & bitfield of 32 boolean flags               & fits in one i32, so it's flattened, in-register                                                                \\
        \hline
        resource-0  & handle-table index to resource             & handle passed in-register + transferred ownership bookkeeping \\
        \hline
    \end{tabular}
    \caption{
        WIT types distinguished by their marshalling behavior.
        Some types are passed in-register for cross-component calls, whether they are spilled and copied over memory depends on flattening thresholds.
        Arguments expressed in $\le16$ core-Wasm values are flattened into registers while results must fit in $\le1$ core-Wasm value to be flattened.
    }
    \label{tab:wit-type}
\end{table*}

Chart (a) shows the latency imposed across the WIT types defined in Table~\ref{tab:wit-type}.
Specialized and tier 3 instrumentation show similar overhead, showing that latency is dominated by the number of component hops.
The overhead is further dominated by marshalling cost.
Tier 1 has no data access, while tier 2 must \emph{lift} payloads into the read-only representation, incurring higher latency.
Avoiding this cross-boundary \emph{lift} makes tier 1 over $2x$ faster for the most expensive WIT type, \texttt{list\textless string\textgreater}.
Chart (b) shows the imposed latency to be approximately linear across all tooling, which corresponds to the copying cost as the \texttt{list} grows.
Note that as the \emph{noop} tool never accesses a payload, these costs are incurred just by the interposition existing.
Optimizations such as \emph{lazy lowering}~\footnote{https://github.com/WebAssembly/component-model/issues/383} would greatly reduce such base costs, only incurring overhead on payload \emph{access}.

\subsubsection{Real Instrumentation Workloads}\label{subsubsec:perf-builtins}

\begin{figure*}
    \includegraphics[width=1.00\linewidth]{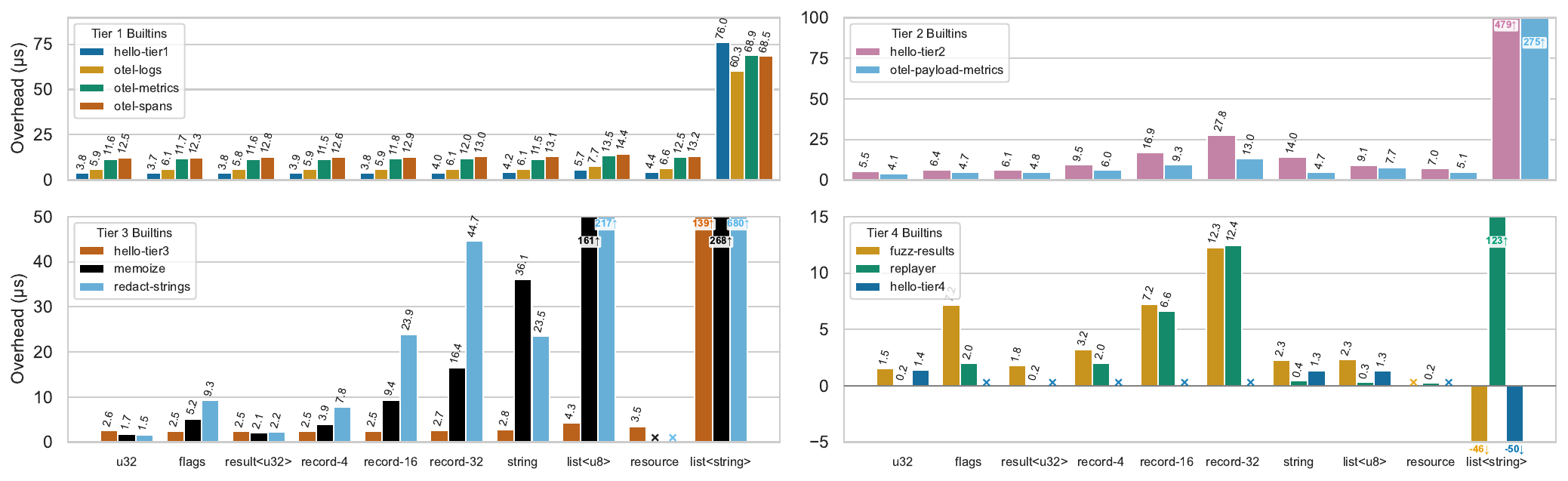}
    \caption{
        The overhead of \ourtool builtins across WIT types.
        Bars show the absolute overhead defined as $p50(builtin) - p50(uninstrumented)$.
        Bars exceeding the y limits are labeled and ``X'' marks types not interposable by the builtin.
    }
    \label{fig:builtins}
\end{figure*}

Figure~\ref{fig:builtins} shows the absolute overhead on the p50 latency imposed by \ourtool builtins (see Table~\ref{tab:builtins}) across WIT types (see Table~\ref{tab:wit-type}).
Note that the y-axis limit differs per chart.
Tier 1 is small and uniform as it does not pass payloads to tools, while others grow with marshalling complexity.
Tier 4 shows that virtualizing the downstream saves time on the expensive \texttt{list\textless string\textgreater} type as it returns an empty list.
While the ABI represents the \texttt{flags} type as a single \texttt{u32}, it is much more expensive than the \texttt{u32} type in Tiers 3--4, because these tiers reify flags as lists of strings.
We also see that any builtin that serializes or deserializes the payload's WAVE representation has a higher cost, including the \texttt{memoize}, \texttt{redact-strings}, \texttt{fuzz-results}, and \texttt{replayer} builtins.
As the \texttt{memoize} builtin skips the downstream request on identical incoming payloads, it is only cheaper if the cost of checking is less than the downstream cost.
The \texttt{redact-strings}, \texttt{fuzz-results}, and \texttt{replayer} builtins all must serialize WAVE representations of a redacted, or synthesized payload into its WIT type.
As \texttt{fuzz-results} and \texttt{replayer} builtins do not require inspecting arguments, they avoid that extra overhead.

\subsubsection{\ourtool vs. State-of-the-Art Frameworks}\label{subsubsec:chosen-fmks}

\begin{table*}
    \renewcommand{\arraystretch}{1.3}
    \newcolumntype{Y}{>{\raggedright\arraybackslash}X}
    \small
    \begin{tabularx}{\textwidth}{@{}l l Y Y Y Y Y@{}}
        \textbf{Architecture} & \textbf{Bind Time}       & \textbf{Placement}                 & \textbf{Granularity}                         & \textbf{Typed Reach}               & \textbf{Language Portability}      & \textbf{Deployment}       \\
        \hline
        \textbf{Sidecar}      & Deploy-time              & Process/network perimeter          & Protocol-level types (L7), raw bytes (L4)    & 1 typed/protocol; $\infty$ untyped & All languages                      & Separate process          \\
        \textbf{eBPF}         & Runtime (dynamic attach) & Kernel/syscall; uprobes w/ symbols & Syscalls, kernel events, network packets     & 0 typed; $\infty$ untyped          & All languages                      & In-kernel + agent process \\
        \textbf{In-language}  & Compile-time             & In-process call sites              & Types exposed by framework                   & N typed, framework-exposed         & 1 (tied to language and framework) & Bundled in app            \\
        \textbf{\ourtool}     & Build-time               & Any edge (perimeter or internal)   & Typed WIT values at arbitrary WIT interfaces & N typed, any interface             & All wasm-targetable languages      & Bundled in app            \\
    \end{tabularx}
    \caption{Comparison of instrumentation framework architectures.}
    \label{tab:feature-matrix}
\end{table*}

To evaluate \ourtool, we categorize other approaches in Table~\ref{tab:feature-matrix} and select representative candidates for empirical evaluation.
In this experiment, we run a simple echo service under increasing load and configure each framework to collect the same metrics (request count and latency).
We use the \texttt{wrk2} workload generator and poll \texttt{docker stats} once per second to collect working-set memory and CPU usage across all containers in the deployment.
We sum memory and CPU usage across all containers, averaged over the measurement window, where $100\%$ means one full CPU core.
For the \textbf{in-language} strategy, we perform an extra function call before the handler logic.
Envoy~\cite{envoy} uses the \textbf{Sidecar} strategy, which routes traffic through a proxy.
Beyla~\cite{beyla} uses \textbf{eBPF} to attach probes in the kernel and user space to observe requests, with a data collector agent that reads from a ring buffer.
\ourtool interposes \emph{before} the handler component which runs in Wasm.

\begin{figure*}
    \includegraphics[width=1.00\linewidth]{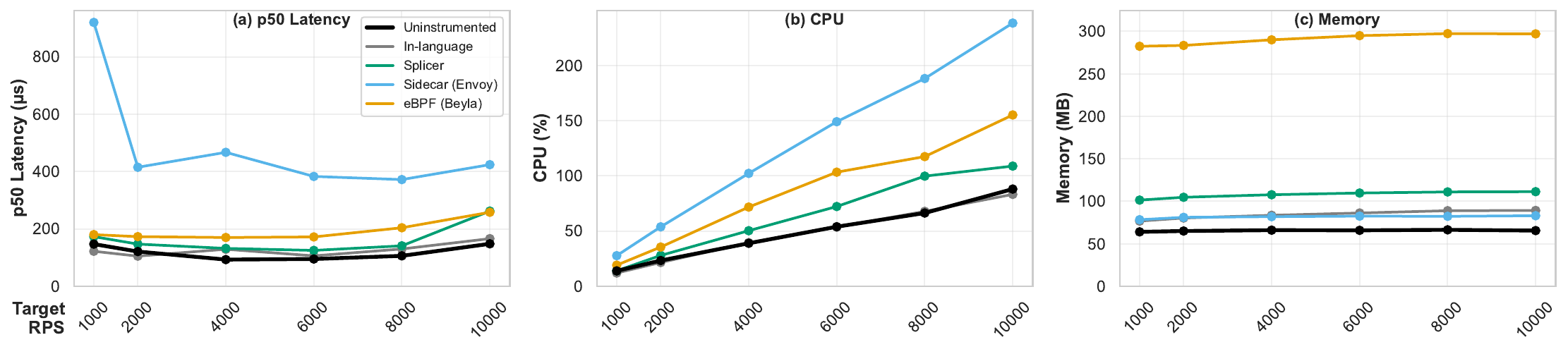}
    \caption{
        The cost of interposing a \emph{noop} tool on a single \texttt{wasi-http} echo service across frameworks, as load increases showing: (a) p50 latency, (b) CPU usage, and (c) container working-set memory (\texttt{docker stats}).
    }
    \label{fig:framework_overhead}
\end{figure*}

Figure~\ref{fig:framework_overhead} shows the p50 (median) latency, CPU, and memory usage of these frameworks.
Our measurements show that the in-language middleware overhead is negligible, barely detectable in latency and tracks CPU percentage closely.
The sidecar imposes high latency due to its added network hop, which is avoided by \ourtool by operating in-process and eBPF by operating inline in the kernel.
Beyla's high memory overhead is due to the extra agent running in the Docker container, which dwarfs the size of the small echo service~\footnote{
In a real deployment, its footprint is negligible as it supports every service running on a single node.
}.
The sidecar likewise adds a separate Envoy proxy process, though a much smaller one.
The service host stores and reuses instances across requests in all deployments.
These instances are larger for in-language middleware and \ourtool as they also contain instrumentation code, with \ourtool embedding it as a whole new component.

\subsubsection{Real-World Application}\label{subsubsec:perf-real-world}
A microservice benchmark does not currently exist for WebAssembly components.
We therefore contribute a component port of the DeathStarBench Hotel Reservation application.
All services are written in Go, compile to Wasm components with TinyGo, and run on the wasmtime engine.
Each service is factored into its core business logic, a \emph{store} component that uses MongoDB, and a \emph{cache} component similar to memcached.
This factoring enables two deployment architectures: one where all components run as separate containers communicating over TCP, or as a single composition communicating over the ABI.

\textbf{Component MSA port with verified equivalence.}
To verify our port's equivalence with the original application, we built a harness that checks the following properties:
\begin{enumerate}
    \item \textbf{Initial state}: The seeded DB is identical at startup.
    \item \textbf{Functionality}: Per-request responses are identical.
    \item \textbf{Final state}: The DB is equivalent after the run.
    \item \textbf{Topology}: Per-request backend service, DB, and cache requests match.
\end{enumerate}
The request corpus comprehensively covers success and failure cases across all endpoints, which is run serially against both applications.
Instrumented executions collect the topology per request and record all responses and DB state, which is used to verify equivalence offline.
We require equivalence across all properties after programmatically accounting for explicitly known architectural differences such as the extra component layer before MongoDB resulting in a different topology from the native Go implementation.

\textbf{Benchmark Corrections.}
To complete our experiments, we corrected a small number of issues in the original benchmark and the component port.
The component port is compiled using TinyGo version 0.41.1 with a minor patch that fixes a GC bug.
We also discovered that the original DeathStarBench had a severe scalability problem where the percentage of writes saturated the \texttt{reservation} service under sustained load resulting in a high number of scans.
We added an index to the DB to correct this in both applications.
Further, to remove confounding overheads from the native Go application, we removed the use of Jaeger, a framework for automatic tracing, and Consul, a service discovery tool
~\footnote{We made all calls to Jaeger no-op and replaced Consul queries with hardcoded IP addresses}
.
These changes ensured that the native Go application had the highest performance possible when compared with the component port.

\begin{figure*}
    \includegraphics[width=1.00\linewidth]{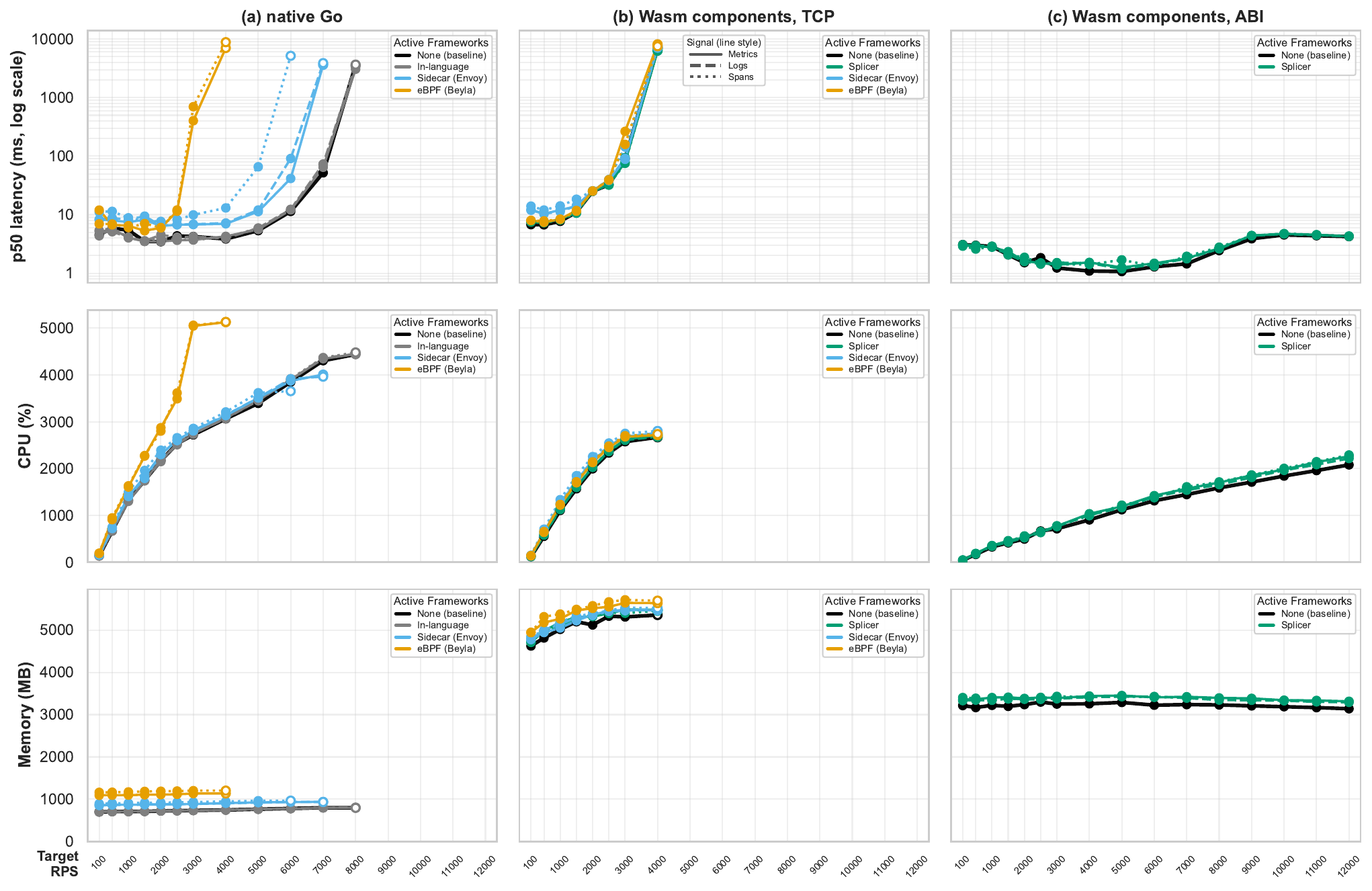}
    \caption{
        The p50 latency, CPU usage, and working-set memory of the DeathStarBench application across frameworks and architectures, varying the requests per second.
        CPU and memory usage are summed across all services in the deployment and averaged over the measurement window.
        Empty circles represent throughput saturation, which is defined as the actual RPS reached being $<90\%$ of the target RPS.
    }
    \label{fig:msa_instr_latency}
\end{figure*}

\textbf{\ourtool on a Real-World MSA.}
We run the Wasm port as individual services hosted in containers communicating over TCP, and as a single composed service communicating over the ABI.
All experiments were run on \texttt{localhost} to keep unpredictable network latencies from affecting TCP measurements.
We use the \texttt{wrk2} workload generator at $75\%$ read, $20\%$ write, and $5\%$ authentication requests.
We compare \ourtool against Go's gRPC middleware, Envoy, and Beyla.
Each instrumented variant targets every service: every container in the distributed deployments, and every component in the ABI composition.
If a deployment saturates (the reached requests per second, RPS, is less than $90\%$ of the target), a hollow marker is used in the plot.

Figure~\ref{fig:msa_instr_latency} shows the p50 latency, CPU usage, and container working-set memory of both the baseline and instrumented applications.
Native Go saturates much later, at 8000 RPS, than Wasm components communicating over TCP, at 2500 RPS.
Wasm components communicating over the ABI do not reach saturation even at 12000 RPS, which holds even for the instrumented deployments.
In-language instrumentation performs best on native as it is a cheap, in-process Go function call on the request handler.
Envoy scales better than Beyla due to a per-request cost.
Measurements indicate that Beyla's CPU usage is significantly higher than Envoy's, yet at saturation the total CPU load sits well below the $96$ available cores.
This CPU usage is concentrated on the \texttt{reservation} service, which consumed roughly 41 cores, nearly $3x$ more than the uninstrumented service; under Envoy, \texttt{reservation}'s usage was essentially unchanged.
Beyla collects request context with \texttt{uprobes}, which execute directly inline on the application's request-handling threads.
These probes fire on every downstream call a service makes, so as load rises, probes compete with the service for its own threads.
This impacts \texttt{reservation} the most, as it performs the most datastore operations and handles about $70\%$ of all requests under our load configuration.
Envoy scales better for this specific application because our load concentrates traffic on a single service, and a sidecar's separate process can spread across idle cores where Beyla's inline probes cannot.
In contrast, the in-process experiment shows that, for \ourtool, adding \emph{any} of the metric, log, and span instrumentation has low overhead in latency, CPU, and memory as load scales beyond 12000 RPS.
This demonstrates that the limiting factor of the application throughput is largely driven by the communication primitive.
While throughput remains steady at high RPS, running such a large composition requires significantly more memory as component instances are stored and reused across requests.

\section{Related Work}\label{sec:related-work}

\textbf{Source-Level instrumentation.}
While a tool can be directly written in or alongside a program's source code (typically manually via SDKs or automatically via AOP~\cite{OrigAOP}), doing so comes with disadvantages.
First, it generates opaque binaries where source and tool code are mixed together, requiring a full rebuild and redeploy of the program to update, add, or remove tools.
As \ourtool operates directly on the Wasm binary, instrumentation can be updated by rerunning the tool without source access.
Second, source-level tools are bound to a specific, target language and must be reimplemented in any new language.
Targeting language-portable WIT interfaces enables \ourtool to port across languages.

\textbf{Rewriting.}
New instructions can be inserted directly into a program by rewriting its source, IR, bytecode~\cite{java-webapp-instr}, or machine code.
\emph{Static rewriting} frameworks, such as Oron~\cite{Oron} and EEL~\cite{EEL}, inject tool code into a program before its execution, producing an instrumented program that ports across execution platforms.
\emph{Dynamic rewriting} frameworks, such as Jalangi~\cite{Jalangi}, Pin~\cite{Pin}, DynamoRIO~\cite{DynamoRIO}, and Frida~\cite{frida}, insert tool code at runtime.
However, this approach incurs an instrumentation binding cost at runtime, slowing program startup.
Hybrid approaches can attempt to reduce such overheads with caching~\cite{HybridRewriting}.
However, as with source-level instrumentation, these approaches produce opaque artifacts that risk changing program behavior inadvertently.
\ourtool performs static rewriting on target compositions, while newly interposed components are ``opaquely'' inserted, they are \emph{sandboxed} in that they share no state with the application, guaranteeing to not affect program behavior.

\textbf{Emulation.}
Emulators and VMs, such as QEMU~\cite{QEMU} and Valgrind~\cite{Valgrind}, enable instrumentation without rewriting.
Rather, they track instrumented program points and invoke tools in a separate state space.
These approaches often make heavy instrumentation prohibitively expensive, though optimizations can achieve high performance~\cite{cross-isa}.
Regardless, they all lock tools into a specific runtime with such capabilities.
\ourtool performs instrumentation by injecting directly into guest bytecode, which ports across any component-supporting Wasm runtime without special handling.

\textbf{Instrumenting at the Program Perimeter.}
Rather than instrumenting the program itself, frameworks that work outside the application and runtime layer can be leveraged to instrument observable program execution \emph{side effects}.
At the process-level abstraction, frameworks such as ptrace~\cite{ptrace}, strace~\cite{strace}, and others~\cite{syscall-no-comp,k23} can observe syscalls and signals.
At the kernel/OS level, frameworks such as DTrace~\cite{DTrace}, eBPF~\cite{eBPF} and Event Tracing for Windows (ETW)~\cite{etw} can observe events like page faults, and scheduling.
Further below, at the hardware/CPU-level, frameworks such as Intel Processor Trace (Intel PT)~\cite{IntelPT} observe control flow.
Some frameworks target webservices specifically.
External network edges can observe client/service requests using API gateways (e.g.\ Kong~\cite{kong} and ContextForge's AI Gateway~\cite{ContextForge}) and Web Application Firewalls (e.g.\ ModSecurity~\cite{modsecurity}).
Within a cluster, service meshes such as Istio~\cite{istio} and Linkerd~\cite{linkerd} can observe and control traffic (L7 HTTP and gRPC) between services.
This is done by intercepting requests via per-workload sidecar proxies (Envoy~\cite{envoy} for Istio and Linkerd Proxy~\cite{linkerd-proxy} for Linkerd).
However, sidecars are known to have disadvantages including high resource consumption, and increased application latency~\cite{sidecar-overheads}.

While these frameworks enable language-portable, high-granularity instrumentation, they are limited by the semantics crossing their specific boundary (bytes and syscalls at the lowest layers, and network protocols at the highest).
These frameworks never achieve the typed, application-level values exchanged inside a workload (except for RPC interposers as this protocol carries high-level types across boundaries).
\ourtool can achieve such high-level semantics through operating on the WIT interface, a communication layer that carries the types itself.
Further, they are constrained to their runtime context (a specific ISA, protocol, etc.), whereas \ourtool is a generic framework that interposes at any component boundary, porting to anywhere Wasm can execute and behind any protocol implemented by the host.

\textbf{Middleware interceptors.}
Middleware exposes typed communication boundaries similar to Wasm's WIT definitions, making them attractive interposition points.
CORBA~\cite{CORBA} supports interposition natively through portable interceptors, binding the interceptor's feature set to the platform's capabilities.
COM~\cite{COM} intercepts without dedicated platform support through binary rewriting.
SDMM~\cite{sdmm} also does this, but generalizes across DDS and gRPC.
These avoid the native-support requirement, but bind to a specific binary format or hardware architecture.
\ourtool is ordinary Wasm component composition, running anywhere Wasm does, and requiring no special runtime support.
This decouples its capabilities from the underlying engine, so its feature set ports across engine versions as it matures.

\textbf{Wasm Instrumentation.}
Wasm \emph{core modules} can be instrumented via static rewriting with frameworks such as BREWasm~\cite{brewasm}, AspectWasm~\cite{AspectWasm}, Walrus~\cite{Walrus}, Wastrumentation~\cite{Wastrumentation}, and Wasabi~\cite{Wasabi}.
Engines with instrumentation support, such as Wizard~\cite{WizardInstr}, can also be used to instrument \emph{core modules} where the tool code exists in a separate state space from the program.
This target enables composable monitoring with high-performance due to intrinsics in the interpreter and JIT tiers.
Whamm~\cite{whamm} is a framework with a DTrace-style DSL that can instrument programs via its engine \emph{or} bytecode rewriting targets.
All of these frameworks can operate on \emph{core modules}, but do not support component-level instrumentation.
A position-paper~\cite{similar-work} proposes a similar architecture to ours; however, it does not provide an automated implementation of the idea, empirically evaluate it, or generalize the applicability of the architecture beyond a DSL for security.
\ourtool enables instrumentation at the component level and exposes a generic framework that can be leveraged across diverse use cases.

\section{Conclusion and Future Work}\label{sec:conclusion}

This work presents \ourtool, a framework that enables cross-language interposition with powerful expression to support diverse use cases.
Through targeting the Wasm Component Model, \ourtool provides instrumentation that: 1) operates on semantically rich types, 2) is sandboxed, and 3) ports seamlessly across polyglot contexts.
We showed it to have acceptable performance directly tied to the performance of the Component Model itself; as the Component Model matures, its optimizations will directly benefit \ourtool.

In future work, we plan to expand the capabilities of \ourtool through supporting any Wasm-targetable language across \emph{all} tiers (tiers 3--4 only support Rust), upcycle the YAML configuration into the full expressive power of a DSL, and expose an interface for ergonomic context propagation between interpositions (like Pivot Tracing's~\cite{PivotTracing} baggage).
While propagation is technically achievable now on tiers 1--2, it is tedious and requires an external agent process.
It should be achievable in \ourtool natively by linking extra metadata-collecting instances into configured tool instances.

\begin{acks}
This work is supported by the WebAssembly Research Center and the NSF Grant Award \#2148301.
We would like to thank Pat Hickey and Oscar Spencer at F5, Bailey Hayes at Cosmonic, Yan Chen at Fastly, and Ewout Verlinde at Ghent University.
\end{acks}

\bibliographystyle{ACM-Reference-Format}
\bibliography{paper}

@inproceedings{java-webapp-instr,
    author = {Cho, Sangwook and Kim, Gyoosik and Cho, Seong-je and Choi, Jongmoo and Park, Minkyu and Han, Sangchul},
    title = {Runtime Input Validation for Java Web Applications using Static Bytecode Instrumentation},
    year = {2016},
    isbn = {9781450344555},
    publisher = {Association for Computing Machinery},
    address = {New York, NY, USA},
    url = {https://doi.org/10.1145/2987386.2987432},
    doi = {10.1145/2987386.2987432},
    booktitle = {Proceedings of the International Conference on Research in Adaptive and Convergent Systems},
    pages = {148–152},
    numpages = {5},
    location = {Odense, Denmark},
    series = {RACS '16}
}

@inproceedings{k23,
    author = {G{\'o}mez Moreno, Jes{\'u}s Mar{\'i}a and Moutafis, Vissarion and Dionysiou, Antreas and Kuipers, Fernando and Smaragdakis, Georgios and Coppens, Bart and Voulimeneas, Alexios},
    title = {Clair Obscur: The Light and Shadow of System Call Interposition – From Pitfalls to Solutions with K23},
    year = {2025},
    isbn = {9798400715549},
    publisher = {Association for Computing Machinery},
    address = {New York, NY, USA},
    url = {https://doi.org/10.1145/3721462.3770772},
    doi = {10.1145/3721462.3770772},
    booktitle = {Proceedings of the 26th International Middleware Conference},
    pages = {241–255},
    numpages = {15},
    location = {Vanderbilt University, Nashville, TN, USA},
    series = {Middleware '25}
}

@INPROCEEDINGS{syscall-no-comp,
    author={Jacobs, Adriaan and Gülmez, Merve and Andries, Alicia and Volckaert, Stijn and Voulimeneas, Alexios},
    booktitle={2024 54th Annual IEEE/IFIP International Conference on Dependable Systems and Networks (DSN)},
    title={System Call Interposition Without Compromise},
    year={2024},
    volume={},
    number={},
    pages={183-194},
    doi={10.1109/DSN58291.2024.00030}}

@inproceedings{cross-isa,
    author = {Cota, Emilio G. and Carloni, Luca P.},
    title = {Cross-ISA machine instrumentation using fast and scalable dynamic binary translation},
    year = {2019},
    isbn = {9781450360203},
    publisher = {Association for Computing Machinery},
    address = {New York, NY, USA},
    url = {https://doi.org/10.1145/3313808.3313811},
    doi = {10.1145/3313808.3313811},
    booktitle = {Proceedings of the 15th ACM SIGPLAN/SIGOPS International Conference on Virtual Execution Environments},
    pages = {74–87},
    numpages = {14},
    location = {Providence, RI, USA},
    series = {VEE 2019}
}

@InProceedings{OrigAOP,
author="Kiczales, Gregor
and Lamping, John
and Mendhekar, Anurag
and Maeda, Chris
and Lopes, Cristina
and Loingtier, Jean-Marc
and Irwin, John",
editor="Ak{\c{s}}it, Mehmet
and Matsuoka, Satoshi",
title="Aspect-oriented programming",
booktitle="ECOOP'97 --- Object-Oriented Programming",
year="1997",
publisher="Springer Berlin Heidelberg",
address="Berlin, Heidelberg",
pages="220--242",
isbn="978-3-540-69127-3"
}

@inbook{IntelPT,
title = "{{Intel 64\textregistered\ and IA-32 Architectures Software Developer’s Manual Volume 3 (3A, 3B, 3C \& 3D): System Programming Guide}}",
year = "2023",
chapter = {33: Intel Processor Trace}
}

@misc{QEMU,
 title={{QEMU}: A generic and open source machine emulator and virtualizer},
 url={http://qemu.org},
 howpublished = {\url{http://qemu.org}},
 author = {Fabrice Bellard},
 year={2020},
 urldate={2023-1-1},
 note={(Accessed 2023-8-07)},
}

@inproceedings{WizardInstr,
    author = {Titzer, Ben L. and Gilbert, Elizabeth and Teo, Bradley Wei Jie and Anand, Yash and Takayama, Kazuyuki and Miller, Heather},
    title = {Flexible Non-intrusive Dynamic Instrumentation for WebAssembly},
    year = {2024},
    isbn = {9798400703867},
    publisher = {Association for Computing Machinery},
    address = {New York, NY, USA},
    url = {https://doi.org/10.1145/3620666.3651338},
    doi = {10.1145/3620666.3651338},
    booktitle = {Proceedings of the 29th ACM International Conference on Architectural Support for Programming Languages and Operating Systems, Volume 3},
    pages = {398–415},
    numpages = {18},
    location = {La Jolla, CA, USA},
    series = {ASPLOS '24}
}

@article{Valgrind,
author = {Nethercote, Nicholas and Seward, Julian},
title = {Valgrind: A Framework for Heavyweight Dynamic Binary Instrumentation},
year = {2007},
issue_date = {June 2007},
publisher = {Association for Computing Machinery},
address = {New York, NY, USA},
volume = {42},
number = {6},
issn = {0362-1340},
url = {https://doi.org/10.1145/1273442.1250746},
doi = {10.1145/1273442.1250746},
journal = {SIGPLAN Not.},
month = {jun},
pages = {89–100},
numpages = {12}
}

@inproceedings{WasmPldi,
author = {Haas, Andreas and Rossberg, Andreas and Schuff, Derek L. and Titzer, Ben L. and Holman, Michael and Gohman, Dan and Wagner, Luke and Zakai, Alon and Bastien, JF},
title = {Bringing the Web up to Speed with {W}eb{A}ssembly},
year = {2017},
isbn = {9781450349888},
publisher = {Association for Computing Machinery},
address = {New York, NY, USA},
url = {https://doi.org/10.1145/3062341.3062363},
doi = {10.1145/3062341.3062363},
booktitle = {Proceedings of the 38th ACM SIGPLAN Conference on Programming Language Design and Implementation},
pages = {185–200},
numpages = {16},
location = {Barcelona, Spain},
series = {PLDI 2017}
}

@misc{CloudflareWasm,
 title={{WebAssembly} on {C}loudflare {W}orkers},
 url={https://blog.cloudflare.com/webassembly-on-cloudflare-workers/},
 howpublished = {\url{https://blog.cloudflare.com/webassembly-on-cloudflare-workers/}},
 publisher = {Cloudflare Blog},
 author = {Varda, Kenton},

 urldate={2018-10-01},
 note={(Accessed 2021-07-06)},
}

@misc{FastlyEdge,
 title={The edge of the multi-cloud},
 url={https://www.fastly.com/cassets/6pk8mg3yh2ee/79dsHLTEfYIMgUwVVllaa4/5e5330572b8f317f72e16696256d8138/WhitePaper-Multi-Cloud.pdf},
 howpublished = {\url{https://www.fastly.com/cassets/6pk8mg3yh2ee/79dsHLTEfYIMgUwVVllaa4/5e5330572b8f317f72e16696256d8138/WhitePaper-Multi-Cloud.pdf}},
 publisher = {Fastly},

 urldate={2020},
 year={2020},
 note={(Accessed 2021-07-06)},
}

@inproceedings{wizard,
author = {Titzer, Ben L. and Gilbert, Elizabeth and Teo, Bradley Wei Jie and Anand, Yash and Takayama, Kazuyuki and Miller, Heather},
title = {Flexible Non-intrusive Dynamic Instrumentation for WebAssembly},
year = {2024},
isbn = {9798400703867},
publisher = {Association for Computing Machinery},
address = {New York, NY, USA},
url = {https://doi.org/10.1145/3620666.3651338},
doi = {10.1145/3620666.3651338},
booktitle = {Proceedings of the 29th ACM International Conference on Architectural Support for Programming Languages and Operating Systems, Volume 3},
pages = {398–415},
numpages = {18},
location = {La Jolla, CA, USA},
series = {ASPLOS '24}
}

@article{Pin,
author = {Luk, Chi-Keung and Cohn, Robert and Muth, Robert and Patil, Harish and Klauser, Artur and Lowney, Geoff and Wallace, Steven and Reddi, Vijay Janapa and Hazelwood, Kim},
title = {Pin: Building Customized Program Analysis Tools with Dynamic Instrumentation},
year = {2005},
issue_date = {June 2005},
publisher = {Association for Computing Machinery},
address = {New York, NY, USA},
volume = {40},
number = {6},
issn = {0362-1340},
url = {https://doi.org/10.1145/1064978.1065034},
doi = {10.1145/1064978.1065034},
journal = {SIGPLAN Not.},
month = jun,
pages = {190–200},
numpages = {11}
}

@book{DTrace,
author = {Gregg, Brendan and Mauro, Jim},
title = {{D}{T}race: Dynamic Tracing in {O}racle {S}olaris, {M}ac {OS} {X} and {F}ree{BSD}},
year = {2011},
isbn = {0132091518},
publisher = {Prentice Hall Press},
address = {USA},
edition = {1st}
}

@article{CORBA,
    author = {Marchetti, Carlo and Verde, Leopoldo},
    year = {2003},
    month = {05},
    pages = {551-579},
    title = {CORBA request portable interceptors: Analysis and applications},
    volume = {15},
    journal = {Concurrency and Computation: Practice and Experience},
    doi = {10.1002/cpe.716}
}

@article{PivotTracing,
    author = {Mace, Jonathan and Roelke, Ryan and Fonseca, Rodrigo},
    title = {Pivot tracing: dynamic causal monitoring for distributed systems},
    year = {2020},
    issue_date = {March 2020},
    publisher = {Association for Computing Machinery},
    address = {New York, NY, USA},
    volume = {63},
    number = {3},
    issn = {0001-0782},
    url = {https://doi.org/10.1145/3378933},
    doi = {10.1145/3378933},
    journal = {Commun. ACM},
    month = feb,
    pages = {94–102},
    numpages = {9}
}

@inproceedings{Oron,
author = {Munsters, A\"{a}ron and Pupo, Angel Luis Scull and Bauwens, Jim and Boix, Elisa Gonzalez},
title = {Oron: Towards a Dynamic Analysis Instrumentation Platform for {A}ssembly{S}cript},
year = {2021},
isbn = {9781450389860},
publisher = {Association for Computing Machinery},
address = {New York, NY, USA},
url = {https://doi.org/10.1145/3464432.3464780},
doi = {10.1145/3464432.3464780},
booktitle = {Companion Proceedings of the 5th International Conference on the Art, Science, and Engineering of Programming},
pages = {6–13},
numpages = {8},
location = {Cambridge, United Kingdom},
series = {Programming '21}
}

@INPROCEEDINGS{AspectWasm,
  author={Rodrigues, João and Barreiros, Jorge},
  booktitle={2022 17th Iberian Conference on Information Systems and Technologies (CISTI)}, 
  title={Aspect-Oriented {W}eb{A}ssembly Transformation}, 
  year={2022},
  volume={},
  number={},
  pages={1-6},
  doi={10.23919/CISTI54924.2022.9820136}}

@InProceedings{WasmIndMach,
author="Nakakaze, Otoya
and Koren, Istv{\'a}n
and Brillowski, Florian
and Klamma, Ralf",
editor="Chbeir, Richard
and Huang, Helen
and Silvestri, Fabrizio
and Manolopoulos, Yannis
and Zhang, Yanchun",
title="Retrofitting Industrial Machines with {W}eb{A}ssembly on the Edge",
booktitle="Web Information Systems Engineering -- WISE 2022",
year="2022",
publisher="Springer International Publishing",
address="Cham",
pages="241--256",
isbn="978-3-031-20891-1"
}

@inproceedings{Wasabi,
author = {Lehmann, Daniel and Pradel, Michael},
title = {Wasabi: A Framework for Dynamically Analyzing {W}eb{A}ssembly},
year = {2019},
isbn = {9781450362405},
publisher = {Association for Computing Machinery},
address = {New York, NY, USA},
url = {https://doi.org/10.1145/3297858.3304068},
doi = {10.1145/3297858.3304068},
booktitle = {Proceedings of the Twenty-Fourth International Conference on Architectural Support for Programming Languages and Operating Systems},
pages = {1045–1058},
numpages = {14},
location = {Providence, RI, USA},
series = {ASPLOS '19}
}

@inproceedings{DeathStarBench,
    author = {Gan, Yu and Zhang, Yanqi and Cheng, Dailun and Shetty, Ankitha and Rathi, Priyal and Katarki, Nayan and Bruno, Ariana and Hu, Justin and Ritchken, Brian and Jackson, Brendon and Hu, Kelvin and Pancholi, Meghna and He, Yuan and Clancy, Brett and Colen, Chris and Wen, Fukang and Leung, Catherine and Wang, Siyuan and Zaruvinsky, Leon and Espinosa, Mateo and Lin, Rick and Liu, Zhongling and Padilla, Jake and Delimitrou, Christina},
    title = {An Open-Source Benchmark Suite for Microservices and Their Hardware-Software Implications for Cloud \& Edge Systems},
    year = {2019},
    isbn = {9781450362405},
    publisher = {Association for Computing Machinery},
    address = {New York, NY, USA},
    url = {https://doi.org/10.1145/3297858.3304013},
    doi = {10.1145/3297858.3304013},
    booktitle = {Proceedings of the Twenty-Fourth International Conference on Architectural Support for Programming Languages and Operating Systems},
    pages = {3–18},
    numpages = {16},
    location = {Providence, RI, USA},
    series = {ASPLOS '19}
}

@inproceedings{HybridRewriting,
author = {Roy, Amitabha and Hand, Steven and Harris, Tim},
title = {Hybrid binary rewriting for memory access instrumentation},
year = {2011},
isbn = {9781450306874},
publisher = {Association for Computing Machinery},
address = {New York, NY, USA},
url = {https://doi.org/10.1145/1952682.1952711},
doi = {10.1145/1952682.1952711},
booktitle = {Proceedings of the 7th ACM SIGPLAN/SIGOPS International Conference on Virtual Execution Environments},
pages = {227–238},
numpages = {12},
location = {Newport Beach, California, USA},
series = {VEE '11}
}

@inproceedings{EEL,
author = {Larus, James R. and Schnarr, Eric},
title = {{EEL}: Machine-Independent Executable Editing},
year = {1995},
isbn = {0897916972},
publisher = {Association for Computing Machinery},
address = {New York, NY, USA},
url = {https://doi.org/10.1145/207110.207163},
doi = {10.1145/207110.207163},
booktitle = {Proceedings of the ACM SIGPLAN 1995 Conference on Programming Language Design and Implementation},
pages = {291–300},
numpages = {10},
location = {La Jolla, California, USA},
series = {PLDI '95}
}

@inproceedings{Jalangi,
author = {Sen, Koushik and Kalasapur, Swaroop and Brutch, Tasneem and Gibbs, Simon},
title = {Jalangi: A Selective Record-Replay and Dynamic Analysis Framework for {J}ava{S}cript},
year = {2013},
isbn = {9781450322379},
publisher = {Association for Computing Machinery},
address = {New York, NY, USA},
url = {https://doi.org/10.1145/2491411.2491447},
doi = {10.1145/2491411.2491447},
booktitle = {Proceedings of the 2013 9th Joint Meeting on Foundations of Software Engineering},
pages = {488–498},
numpages = {11},
location = {Saint Petersburg, Russia},
series = {ESEC/FSE 2013}
}

@INPROCEEDINGS{DynamoRIO,
  title      = "An infrastructure for adaptive dynamic optimization",
  booktitle  = "International Symposium on Code Generation and Optimization,
                2003. {CGO} 2003",
  author     = "Bruening, D and Garnett, T and Amarasinghe, S",
  publisher  = "IEEE Comput. Soc",
  year       =  2003,
  conference = "International Symposium on Code Generation and Optimization.
                CGO 2003",
  location   = "San Francisco, CA, USA"
}

@misc{Walrus,
 title={Walrus: A {W}eb{A}ssembly transformation library},
 url={https://github.com/rustwasm/walrus},
 howpublished = {\url{https://github.com/rustwasm/walrus}},
 publisher = {GitHub},

 urldate={2023},
 year={2023},
}

@inproceedings{brewasm,
author = {Cao, Shangtong and He, Ningyu and Guo, Yao and Wang, Haoyu},
title = {BREWasm: A General Static Binary Rewriting Framework for WebAssembly},
year = {2023},
isbn = {978-3-031-44244-5},
publisher = {Springer-Verlag},
address = {Berlin, Heidelberg},
url = {https://doi.org/10.1007/978-3-031-44245-2_8},
doi = {10.1007/978-3-031-44245-2_8},
booktitle = {Static Analysis: 30th International Symposium, SAS 2023, Cascais, Portugal, October 22–24, 2023, Proceedings},
pages = {139–163},
numpages = {25},
location = {Lisbon, Portugal}
}

@inproceedings{eBPF,
    author = {McCanne, Steven and Jacobson, Van},
    title = {The BSD packet filter: a new architecture for user-level packet capture},
    year = {1993},
    publisher = {USENIX Association},
    address = {USA},
    booktitle = {Proceedings of the USENIX Winter 1993 Conference Proceedings on USENIX Winter 1993 Conference Proceedings},
    pages = {2},
    numpages = {1},
    location = {San Diego, California},
    series = {USENIX'93}
}

@misc{etw,
    author = {{Microsoft}},
    title = {Event Tracing for {Windows}},
    url = {https://learn.microsoft.com/en-us/windows/win32/etw/event-tracing-portal},
    year = {2024},
    note = {Accessed: 2026-04-28},
}

@misc{kong,
    author = {{Kong Inc.}},
    title = {Kong: The {Cloud} Connectivity Company},
    url = {https://konghq.com/},
    year = {2024},
    note = {Accessed: 2026-04-28},
}

@misc{modsecurity,
    author = {{OWASP ModSecurity Project}},
    title = {ModSecurity: Open Source Web Application Firewall},
    url = {https://github.com/owasp-modsecurity/ModSecurity},
    year = {2024},
    note = {Accessed: 2026-04-28},
}

@misc{ptrace,
    author = {{Linux Manual}},
    title = {{ptrace(2)}: process trace --- {Linux} {man-pages}},
    url = {https://man7.org/linux/man-pages/man2/ptrace.2.html},
    year = {1999},
    note = {Accessed: 2026-04-28},
}

@misc{strace,
    author       = {{strace project}},
    title        = {strace: linux syscall tracer},
    url = {https://strace.io/},
    year         = {2024},
    note         = {Accessed: 2026-04-28},
}

@misc{istio,
    author = {{The Istio Authors}},
    title = {Istio: Connect, Secure, Control, and Observe Services},
    url = {https://istio.io/},
    year = {2025},
    note = {Accessed: 2026-04-28},
}

@misc{linkerd,
    author       = {{The Linkerd Authors}},
    title        = {Linkerd: The world's lightest, fastest service mesh},
    url = {https://linkerd.io/},
    year         = {2026},
    note         = {Accessed: 2026-04-28},
}

@misc{envoy,
    author       = {{The Envoy Authors}},
    title        = {Envoy Proxy},
    url = {https://www.envoyproxy.io/},
    year         = {2024},
    note         = {Accessed: 2026-04-28},
}

@misc{beyla,
    author       = {{Grafana Labs}},
    title        = {eBPF application auto-instrumentation},
    url = {https://grafana.com/docs/beyla/latest/},
    year         = {2026},
    note         = {Accessed: 2026-09-11},
}

@inproceedings{TheMM,
    title={The Mystery Machine: End-to-end Performance Analysis of Large-scale Internet Services},
    author={Michael Chow and David Meisner and Jason Flinn and Daniel Peek and Thomas F. Wenisch},
    booktitle={USENIX Symposium on Operating Systems Design and Implementation},
    year={2014},
    url={https://api.semanticscholar.org/CorpusID:5661499}
}

@inproceedings{Dapper,
    title={Dapper, a Large-Scale Distributed Systems Tracing Infrastructure},
    author={Benjamin H. Sigelman and Luiz Andr{\'e} Barroso and Michael Burrows and Patrick Stephenson and Manoj Plakal and Donald Beaver and Saul Jaspan and Chandan Kumar Shanbhag},
    year={2010},
    url={https://api.semanticscholar.org/CorpusID:14271421}
}

@inproceedings{xtrace,
    author = {Fonseca, Rodrigo and Porter, George and Katz, Randy and Shenker, Scott and Stoica, Ion},
    year = {2007},
    month = {01},
    pages = {},
    title = {X-Trace: A Pervasive Network Tracing Framework.}
}

@inproceedings{Gremlin,
    author={Heorhiadi, Victor and Rajagopalan, Shriram and Jamjoom, Hani and Reiter, Michael K. and Sekar, Vyas},
    booktitle={2016 IEEE 36th International Conference on Distributed Computing Systems (ICDCS)},
    title={Gremlin: Systematic Resilience Testing of Microservices},
    year={2016},
    volume={},
    number={},
    pages={57-66},
    doi={10.1109/ICDCS.2016.11}}

@inproceedings{Filibuster,
    author = {Meiklejohn, Christopher S. and Estrada, Andrea and Song, Yiwen and Miller, Heather and Padhye, Rohan},
    title = {Service-Level Fault Injection Testing},
    year = {2021},
    isbn = {9781450386388},
    publisher = {Association for Computing Machinery},
    address = {New York, NY, USA},
    url = {https://doi.org/10.1145/3472883.3487005},
    doi = {10.1145/3472883.3487005},
    booktitle = {Proceedings of the ACM Symposium on Cloud Computing},
    pages = {388–402},
    numpages = {15},
    location = {Seattle, WA, USA},
    series = {SoCC '21}
}

@inproceedings{replay,
    author = {De Bosschere, Koen and Kergommeaux, Jacques},
    year = {2000},
    month = {01},
    pages = {},
    title = {Execution replay and debugging.}
}

@INPROCEEDINGS{mocking,
    author={Mostafa, Shaikh and Wang, Xiaoyin},
    booktitle={2014 14th International Conference on Quality Software},
    title={An Empirical Study on the Usage of Mocking Frameworks in Software Testing},
    year={2014},
    volume={},
    number={},
    pages={127-132},
    doi={10.1109/QSIC.2014.19}}

@InProceedings{data-prov,
    author="Buneman, Peter
and Khanna, Sanjeev
and Wang-Chiew, Tan",
    editor="Van den Bussche, Jan
and Vianu, Victor",
    title="Why and Where: A Characterization of Data Provenance",
    booktitle="Database Theory --- ICDT 2001",
    year="2001",
    publisher="Springer Berlin Heidelberg",
    address="Berlin, Heidelberg",
    pages="316--330",
    isbn="978-3-540-44503-6"
}

@book{resilience,
    author = {Nygard, Michael},
    title = {Release It! Design and Deploy Production-Ready Software},
    year = {2007},
    isbn = {0978739213},
    publisher = {Pragmatic Bookshelf}
}

@misc{linkerd-proxy,
    author       = {{The Linkerd Authors}},
    title        = {linkerd2-proxy: A purpose-built proxy for the {Linkerd} service mesh},
    url = {https://github.com/linkerd/linkerd2-proxy},
    year         = {2024},
    note         = {Accessed: 2026-04-28},
}

@inproceedings{sidecar-overheads,
    author = {Zhu, Xiangfeng and She, Guozhen and Xue, Bowen and Zhang, Yu and Zhang, Yongsu and Zou, Xuan Kelvin and Duan, XiongChun and He, Peng and Krishnamurthy, Arvind and Lentz, Matthew and Zhuo, Danyang and Mahajan, Ratul},
    title = {Dissecting Overheads of Service Mesh Sidecars},
    year = {2023},
    isbn = {9798400703874},
    publisher = {Association for Computing Machinery},
    address = {New York, NY, USA},
    url = {https://doi.org/10.1145/3620678.3624652},
    doi = {10.1145/3620678.3624652},
    booktitle = {Proceedings of the 2023 ACM Symposium on Cloud Computing},
    pages = {142–157},
    numpages = {16},
    location = {Santa Cruz, CA, USA},
    series = {SoCC '23}
}

@misc{frida,
    author       = {{Frida}},
    title        = {Frida: A world-class dynamic instrumentation toolkit},
    url = {https://frida.re/},
    year         = {2024},
    note         = {Accessed: 2026-04-28},
}

@misc{ContextForge,
    author       = {{ContextForge}},
    title        = {Plugin Framework Specification},
    url = {https://ibm.github.io/mcp-context-forge/1.0.0-RC3/architecture/plugins/},
    year         = {2026},
    note         = {Accessed: 2026-09-21},
}

@article{sdmm,
    title = {A generalized approach to real-time, non-intrusive instrumentation and monitoring of standards-based distributed middleware},
    journal = {Journal of Systems Architecture},
    volume = {117},
    pages = {102181},
    year = {2021},
    issn = {1383-7621},
    doi = {https://doi.org/10.1016/j.sysarc.2021.102181},
    url = {https://www.sciencedirect.com/science/article/pii/S1383762121001296},
    author = {Nyalia Lui and James H. Hill}
}

@inproceedings{COM,
    author = {Hunt, Galen C. and Scott, Michael L.},
    title = {Intercepting and instrumenting COM applications},
    year = {1999},
    publisher = {USENIX Association},
    address = {USA},
    booktitle = {Proceedings of the 5th Conference on USENIX Conference on Object-Oriented Technologies \& Systems - Volume 5},
    pages = {4},
    numpages = {1},
    location = {San Diego, California},
    series = {COOTS'99}
}

@inproceedings{WasmSD-Edge,
    author = {Wang, Rouyi and Zhou, Zhi and Chen, Xu},
    title = {Demo: WasmSD-Edge: A Lightweight Edge Stable Diffusion Image Generation Framework Based on WebAssembly},
    year = {2025},
    isbn = {9798400711299},
    publisher = {Association for Computing Machinery},
    address = {New York, NY, USA},
    url = {https://doi.org/10.1145/3680207.3765593},
    doi = {10.1145/3680207.3765593},
    booktitle = {Proceedings of the 31st Annual International Conference on Mobile Computing and Networking},
    pages = {1210–1212},
    numpages = {3},
    location = {Kerry Hotel, Hong Kong, Hong Kong, China},
    series = {ACM MOBICOM '25}
}

@misc{wasmcloud-cncf,
    author       = {{Cloud Native Computing Foundation}},
    title        = {wasmCloud --- {CNCF} Project},
    url = {https://www.cncf.io/projects/wasmcloud/},
    year         = {2024},
    note         = {Accessed: 2026-04-28},
}

@inproceedings{wasm-iot,
    author={Zyrianoff, Ivan and Montori, Federico and Trotta, Angelo and Sciullo, Luca and Gigli, Lorenzo and Kamienski, Carlos and Di Felice, Marco},
    booktitle={2025 IEEE 22nd Consumer Communications \& Networking Conference (CCNC)},
    title={A Location-Aware WebAssembly-Based Software Update Framework for IoT End Devices},
    year={2025},
    volume={},
    number={},
    pages={1-4},
    doi={10.1109/CCNC54725.2025.10976056}}

@inproceedings{wpu,
    author={Kim, Jinyeol and Lee, Chaebin and Oh, Jongwon and Park, Joungmin and Lee, Seung Eun},
    booktitle={2026 31st Asia and South Pacific Design Automation Conference (ASP-DAC)},
    title={WPU: A Pipelined WebAssembly Processing Unit for Embedded IoT Systems},
    year={2026},
    volume={},
    number={},
    pages={1357-1363},
    doi={10.1109/ASP-DAC66049.2026.11420760}}

@misc{component-model,
    author       = {{WebAssembly Community Group}},
    title        = {{WebAssembly} Component Model},
    url = {https://github.com/WebAssembly/component-model},
    year         = {2024},
    note         = {Accessed: 2026-04-28},
}

@misc{wit,
    author       = {{WebAssembly Community Group}},
    title        = {{WIT}: {WebAssembly} Interface Type},
    url = {https://github.com/WebAssembly/component-model/blob/main/design/mvp/WIT.md},
    year         = {2024},
    note         = {Accessed: 2026-04-28},
}

@misc{wac,
    author       = {{Bytecode Alliance}},
    title        = {{WAC}: {WebAssembly} Compositions},
    url = {https://github.com/bytecodealliance/wac},
    year         = {2024},
    note         = {Accessed: 2026-04-28},
}

@misc{canon-abi,
    author       = {{WebAssembly Community Group}},
    title        = {Canonical ABI Explainer},
    url = {https://github.com/WebAssembly/component-model/blob/main/design/mvp/CanonicalABI.md},
    year         = {2026},
    note         = {Accessed: 2026-04-28},
}

@misc{wirm,
    author       = {{wirm}},
    title        = {WebAssembly IR Manipulation Library for the Component Model},
    url = {https://github.com/composablesys/wirm},
    year         = {2026},
    note         = {Accessed: 2026-04-28},
}

@article{whamm,
    author = {Gilbert, Elizabeth and Schneider, Matthew and An, Zixi and Thalanki, Suhas and Bowman, Wavid and Bai, Alexander Y. and Titzer, Ben L. and Miller, Heather},
    title = {Debugging WebAssembly? Put Some Whamm on It!},
    year = {2025},
    issue_date = {October 2025},
    publisher = {Association for Computing Machinery},
    address = {New York, NY, USA},
    volume = {9},
    number = {OOPSLA2},
    url = {https://doi.org/10.1145/3763124},
    doi = {10.1145/3763124},
    journal = {Proc. ACM Program. Lang.},
    month = oct,
    articleno = {346},
    numpages = {29}
}

@inproceedings{similar-work,
    author = {Blaak, Olav and Van Cutsem, Tom},
    title = {Towards Least-Privilege WebAssembly Applications: Transparent Interposition for WebAssembly Components},
    year = {2026},
    isbn = {9798400725555},
    publisher = {Association for Computing Machinery},
    address = {New York, NY, USA},
    url = {https://doi.org/10.1145/3801119.3801125},
    doi = {10.1145/3801119.3801125},
    booktitle = {Companion Proceedings of the 10th International Conference on the Art, Science, and Engineering of Programming},
    pages = {14–17},
    numpages = {4},
    location = {},
    series = {‹Programming› Companion '26}
}

@inproceedings{Wastrumentation,
    author = {Munsters, Aäron and Scull Pupo, Angel Luis and Gonzalez Boix, Elisa},
    title = {Wastrumentation: Portable WebAssembly Dynamic Analysis with Support for Intercession},
    year = {2025},
    isbn = {978-3-95977-373-7},
    issn = {1868-8969},
    volume = {333},
    editor = {Aldrich, Jonathan and Silva, Alexandra},
    publisher = {Schloss Dagstuhl -- Leibniz-Zentrum f{\"u}r Informatik},
    address = {Dagstuhl, Germany},
    url = {https://drops.dagstuhl.de/entities/document/10.4230/LIPIcs.ECOOP.2025.23},
    urn = {urn:nbn:de:0030-drops-233153},
    doi = {10.4230/LIPIcs.ECOOP.2025.23},
    booktitle = {39th European Conference on Object-Oriented Programming (ECOOP 2025)},
    pages = {23:1--23:29},
    series = {Leibniz International Proceedings in Informatics (LIPIcs)}
}

\end{document}